\documentclass[useAMS, usenatbib]{mn2e}

\usepackage{url}
\usepackage{deluxetable}
\usepackage{comment}
\usepackage{graphicx}
\usepackage{color}

\newcommand{\mtwenty}{$M_{20}\,$}

\newcommand{\hinv}{h^{-1}}
\newcommand{\lcdm}{\Lambda\rm{CDM}}
\newcommand{\gm}{$Gini$-$M_{20}\,$}
\newcommand{\Mhalo}{M_{\rm halo}}
\newcommand{\Mstar}{M_\star}
\newcommand{\Msun}{M_\odot}
\newcommand{\sphgr}{{\small SPHGR}\,}

\title{Identifying mergers using non-parametric morphological classification at high redshifts}

\author[Thompson et al.]{Robert Thompson$^{1,2,3}$, Romeel Dav{\'e}$^{2,4,5}$, Shuiyao Huang$^{6}$, Neal Katz$^{6}$
\\$^1$ NCSA, University of Illinois, Urbana-Champaign, IL 61820
\\$^2$ University of the Western Cape, Bellville, Cape Town 7535, South Africa
\\$^3$ Astronomy Department, University of Arizona, Tucson, AZ 85721, USA
\\$^4$ South African Astronomical Observatories, Observatory, Cape Town 7925, South Africa
\\$^5$ African Institute for Mathematical Sciences, Muizenberg, Cape Town 7945, South Africa
\\$^6$Astronomy Department, University of Massachusetts, Amherst, MA 01003, USA
}

\begin{document}
\maketitle 


 \begin{abstract}
We investigate the time evolution of non-parametric morphological
quantities and their relationship to major mergers between $4\geq
z \geq 2$ in high-resolution cosmological zoom simulations of disk
galaxies that implement kinetic wind feedback, $H_2$-based star formation, and
minimal ISM pressurisation. We show that the resulting galaxies broadly
match basic observed physical properties of $z\sim 2$ objects.  We measure the
galaxies' concentrations ($C$), asymmetries ($A$), and $Gini$ ($G$) and
$M_{20}$ coefficients, and correlate these with major merger
events identified from the mass growth history.  We find that high
values of asymmetry provide the best indicator for identifying
major mergers of $>1:4$ mass ratio within our sample, with
\gm merger classification only as effective for face-on systems
and much less effective for edge-on or randomly-oriented galaxies.
The canonical asymmetry cut of $A\geq0.35$, however, is only able
to correctly identify major mergers $\sim 10\%$ of the time, while
a higher cut of $A\geq 0.8$ more efficiently picks out mergers at
this epoch.  We further examine the temporal correlation between
morphological statistics and mergers, and show that for randomly-oriented
galaxies, half the galaxies with $A\geq0.8$ undergo a merger within
$\pm0.2\,{\rm Gyr}$, whereas \gm identification only identifies
about a third correctly.  The fraction improves further using $A\geq
1.5$, but about the half the mergers are missed by this stringent
cut.
\end{abstract}



\section{Introduction}
\label{sec:introduction}

A key to understanding galaxy evolution is the ability to interpret
their morphologies.  Galaxies are believed to grow and evolve via
in-situ star formation, major and minor mergers, and gas accretion
from the intergalactic and circumgalactic medium.
Understanding a galaxy's morphology
provides an important avenue for understanding how these processes
are responsible for establishing the observed structural features
of the galaxy population across cosmic time \citep{Conselice14},
such as the origin and emergence of the present-day Hubble sequence.

Visual classification along the Hubble sequence has a long history
as a valuable descriptor of galaxy properties.  Morphology is
observed to correlate with many other properties such as colour,
star formation rate, surface brightness, and gas content,  However,
visual classification remains ultimately subjective.  Hence there
has been much interest in developing quantitative classification schemes.
Surface brightness profiles \citep{deVaucouleurs48,Sersic63} provide such
an avenue for quantitatively characterising the light distribution.
Such parametric characterisations must nonetheless make assumptions about 
the underlying functional form that may not hold for all stages of 
galaxy formation.  For instance, they are less well-defined in the
case of merging or other irregular galaxies.

More recently, non-parametric methods have gained popularity for
quantifying morphology.  In this case, the pixelised image is
directly characterised into a mathematical form.  Some common methods
in use today are the Concentration-Asymmetry-Clumpiness (CAS) system
\citep{Conselice03}, the $Gini$ ($G$) coefficient, and the second-order
moment of the brightest 20\% of a galaxy's pixels (\mtwenty)
\citep{Lotz04}.
Since these non-parametric quantities do not
make any assumptions about the underlying form of galactic structure,
they are considered a less biased tool to identify objects that
are undergoing key transformative events such as mergers.

Major mergers are thought to play an important role in the evolution
of galaxies.  In addition to transforming the morphology towards
earlier type \citep[e.g.][]{Toomre72,Cox08}, they are believed to
be responsible for generating bursts of star formation~\citep{Mihos96,Narayanan10,Narayanan10b}
as well as stimulating rapid growth of the central supermassive
black hole~\citep{Sanders96}.  While it is now clear despite the
resulting starbursts, such mergers do not dominate the cosmic star
formation at any observable epoch~\citep[e.g.][]{Rodighiero11},
nonetheless their importance to galaxy transformation and black
hole growth makes the identification of mergers a key challenge in
forming a complete picture for the emergence of the galaxy population.

While non-parametric morphologies offer a model-independent approach
to characterising mergers, they must be calibrated against other
schemes, typically visual classification, to interpret
them.  This can be most easily done at low redshifts, where the
galaxies can be well-resolved and are thus straightforward to
classify.  In this way, CAS and \gm have been shown to isolate
mergers in particular regions of parameter space.  Using this, CAS
has been used to identify major merger candidates in numerous galaxy
surveys \citep[e.g.][]{Abraham96,Conselice03,Cassata05,Menanteau06}.
\citet{Lotz04} argued that \gm is more effective at identifying
late-stage mergers and classifying low-redshift galaxies.

As galaxy surveys are extended to higher redshifts, such parametrisations
continue to be utilised to classify mergers and morphologies.  With
surveys such as CANDELS~\citep{Grogin11,Koekemoer11} using {\it Hubble},
the morphology of distant galaxies can be resolved at rest-optical
wavelengths, which if interpreted properly could provide important
information towards understanding the evolution of galaxies during
the peak epoch of cosmic star formation.  However, visual morphologies
of high-redshift galaxies do not clearly follow the Hubble
sequence classification seen today.  For example, at $z\sim 2$
disks tend to be thicker, more turbulent, and
clumpier~\citep[e.g.][]{ForsterSchreiber09,Elmegreen09,Genzel11},
ellipticals are substantially more compact~\citep{vanDokkum14}, and
the fraction of irregulars increases substantially compared to
today.  It is thus not clear how well non-parametric
morphologies calibrated at low-$z$ will apply to such distant
galaxies, and hence whether quantitative morphology can be effectively
utilised to characterise the high-$z$ galaxy population.

Cosmological hydrodynamical simulations provide a promising avenue
to better understand how quantitative morphologies relate to key
evolutionary events in galaxies such as mergers.  With improving
spatial resolution and a better understanding of the basic physical
ingredients required to broadly reproduce observed
galaxies~\citep{Somerville15}, they have emerged as a valuable
complementary tool to elucidate the physical processes that shape
galaxies.  With sophisticated analysis tools, simulated galaxies
can be processed into mock observations that may be analysed in an
analogous fashion to real data, with the advantage that the true
underlying distribution of the mass, light, and metals is known.

\citet{Lotz08} investigated both CAS and \gm statistics for a number
mock observed non-cosmological, isolated, merging systems in
hydrodynamic simulations.  It was found that \gm identified mergers
at their first passage and when the system finally coalesced, but
did not consistently identify mergers at intermediate times.
\citet{Lotz08,Lotz10sims,Lotz10simsgas} further went on to characterise
the time-scales for structural mergers for $A$ and \gm to be of
similar order.  These simulations were used to calibrate the \gm
merger identification relation for observed galaxies out to $z>3$.

A different approach was taken by \citet{Cibinel15}, who classified
mergers using resolved stellar maps as opposed to single-band images.
They found that a combination of $A$ and M$_{20}$ measured on the
stellar mass maps of $\sim100$ HUDF galaxies provided the most
accurate distinction between mergers and isolated clumpy galaxies.
They found that merger samples identified on the basis of $A$ from
a single $H-band$ image resulted in a contamination as high as
$\sim50\%$ from clumpy galaxies, and argue that their method reduces
this contamination.

While informative, isolated galaxy simulations such as those used
in these studies do not capture the cosmological context of galaxy
formation.  They lack crucial ingredients such as the frequent
infall of smaller satellites (i.e. minor mergers), and important
avenues for galaxy growth by smooth accretion \citep[which is
believed to be the dominant fuelling mode;][]{Keres05}.  Given that
the merger and accretion rates are much higher in the past than
today, morphology could be strongly impacted by such processes.

More recently, the focus has shifted towards using cosmological
zoom-in simulations, in which an individual galaxy and its environs
are evolved at much higher resolution relative to the entire volume.
This retains the full cosmological context while making feasible
the high resolution needed to study internal structural processes
within galaxies such as violent disk instabilities~\citep{Dekel09}.
\citet{Snyder15} examined the \gm statistics for a number of high-resolution
zoom-in galaxies at $3>z>1$.  They found that \gm was more sensitive to
early stages of mergers ($t<0$).  Overall they concluded that 
structural evolution is neither universal nor monotonic, and that \gm were 
largely sensitive to viewing angle, star formation, and mergers.

In this paper we examine a number of galaxies at $4~\geq~z~\geq~2$
within high-resolution cosmological smoothed particle hydrodynamic
zoom-in simulations.  We mock observe our galaxy sample as as if
they were detected in the $H-band$ CANDLES survey with {\it HST}
and investigate their $C$, $A$, $G$, and \mtwenty \,values as a
function of their mass histories.  Our primary aim is to determine
the optimal approach for quantifying the merger history of galaxies
using non-parametric morphological statistics.  The hope is that
this will provide a better understanding of how such statistics can
best be used to interpret the evolution of galaxies at early epochs.

This paper is organised as follows:  in \S\ref{sec:methods}
we describe our simulations and the methods.  \S\ref{sec:globals}
examines the global properties of our galaxy sample.
\S\ref{sec:morphstats} looks at the morphological statistics,
while \S\ref{sec:morphstatshist} investigates the morphological
statistics of each galaxy as a function of cosmic time.  Finally
in \S\ref{sec:conclusions} we summarise our conclusions.


\section{Methods}
\label{sec:methods}


\subsection{Simulation Code}
\label{sec:sims}

Our simulations are evolved with an extended version of the
{\small{GADGET-3}} cosmological SPH code \citep{Springel05,Huang15}.
It includes cooling processes using the primordial abundances
described in \citet{Katz96}, with additional cooling from metal
lines assuming photo-ionisation equilibrium from \citet{Wiersma09}.
We account for metal enrichment from Type II supernovae (SNe), Type
Ia SNe, AGB stars, and we track four elements (C,O,Si,Fe) individually
as described in \citet{Oppenheimer08}.  Galactic outflows are
implemented using the hybrid energy/momentum-driven wind (ezw) model
fully detailed in \citet{Dave13} and \citet{Ford15}.

Compared to our previous code described in \citet{Oppenheimer08},
we now use the `pressure-entropy' SPH algorithm that better handles
issues of two-phase instabilities when compared to the standard
`density-entropy' SPH formulations \citep[see][for further
details]{Saitoh13,Hopkins13}.  
Following \citet{Hopkins13} we use a standard quintic spline kernel
with 128 neighbours and adopt the \citet{MM97} artificial viscosity
along with a \citet{Balsara89} switch.
We also implement a time-step
limiter \citep{Saitoh09,Durier12} that improves the accuracy of
the time integration scheme in situations where there are sudden
changes to a particle's internal energy.

To suppress artificial fragmentation within the interstellar medium,
we prevent gas particles from cooling below their effective Jeans
temperature \citep{KatzGunn91,Schaye08,Robertson08} by ensuring that the
simulation is always resolving at least one Jeans mass within a
particle's smoothing length.  This translates to a threshold number
density for pressurisation of
\begin{equation}
\rho_{th} = \frac{3}{4\pi}\left(\frac{5kT}{G}\right)^3\left(\frac{1}{N_{ngb}\,m_{\rm gas}}\right)^2,
\label{eq:JMT}
\end{equation}
where $\rm{T}=10^4\rm{K}$, $m_H$ is the mass of a hydrogen atom,
$k$ is the Boltzmann constant, $N_{ngb}$ is the number of neighbours
within an SPH kernel, and $m_{\rm gas}$ is the mass of an individual gas
particle within the simulation (one Jeans mass here is $N_{ngb}m_{\rm gas}$).  
Particles with densities above $\rho_{th}$ and are considered to be 
multi-phase, and a Jeans temperature floor is imposed  
\begin{equation} 
T_{\rm Jeans} = 10^4 (\rho/\rho_{th})^{4/3},
\label{eq:Tjeans} 
\end{equation}
where $\rho$ is the gas particle's density.  In low-resolution
simulations, $\rho_{th}$ can drop below the nominal star formation
threshold of $n=0.13\,cm^{-3}$, in which case we set it to this
value; however, in the zoom simulations presented in this work,
this does not occur.
The net effect is analogous to adding ISM pressure \citep{Springel03},
but rather than directly pressurising the gas, we prevent the gas from cooling
and thus fragmenting below the Jeans scale.  The net effect of this
implementation is in practice quite similar to that of \citet{Schaye08},
who force gas particles to lie on the $T\propto \rho^{4/3}$ relation,
except that rather than using a fixed density threshold (as they use), we adjust
our density threshold $\rho_{th}$ to naturally increase as the resolution of
the simulation increases.  This minimises the
amount of artificial pressure introduced into the ISM, which 
more accurately captures the internal structure of galaxies while 
suppressing fragmentation on scales that we cannot resolve.

To model star formation, we employ a molecular gas-based prescription.
The molecular content ($f_{\rm{H_2}}$) of each gas particle is
calculated following the analytic model of \citet[][hereafter
KMT]{Krumholz08,Krumholz09} and \citet{McKee10};  details can be found in
\citet{Thompson14}.  We stochastically form stars from the gas following
a \citet{Schmidt59} Law as in
\begin{equation}
\dot\rho_* = \epsilon_* f_{\rm H_2}{\rho\over t_{\rm ff}}.
\end{equation}
The efficiency of star formation $\epsilon_*$ is set to 1\% per
local free-fall time $t_{\rm ff}$ \citep{Krumholz07,Lada10}.  This
model allows us to regulate star formation using the local abundance
of H$_2$ rather than the total gas density, which confines star
formation to the densest peaks in the ISM as is typically seen in
real galaxies, thereby preserving the ISM structure as much as the
numerical resolution allows.  This also introduces a metallicity
dependence into the star formation efficiency, since the KMT model
of molecular gas formation relies on the presence of dust, which
may be important for suppressing early star
formation~\citep{Tassis12,KrumholzDekel12}.


\subsection{Initial Conditions}
\label{sec:ics}

Initial conditions are generated using the {\sc Music} code
\citep{MUSIC} assuming a flat $\lcdm$ cosmology.  Cosmological
parameters are chosen to be consistent with constraints from the
\citet{Planck13} results, namely $\Omega_{m}=0.3,
\Omega_{\Lambda}=0.7,H_0=70,\sigma_8=0.8,n_s=0.96$.

Our target halos are selected at $z=2$ from a low-resolution 
hydrodynamic simulation consisting of $256^3$ dark-matter 
and gas particles in a $(16\hinv$Mpc$)^3$ volume with an 
effective comoving spatial resolution
of $\epsilon_{grav}=1.25\,\hinv$ kpc, which is $417\,\hinv$ pc at $z=2$.
We populate each target halo with
higher resolution particles at $z=249$, with the size of each higher resolution
region chosen to be 2.5 times the maximum radius of the original
halo from the low-resolution 
hydrodynamic simulation.  The majority of halos in
our sample are simulated with a single additional level of refinement
($\epsilon_{grav}=625\,\hinv$ pc or $208\,\hinv$ pc at $z=2$), while the two
lowest mass halos are simulated with two additional levels of refinement
($\epsilon_{grav}=313\,\hinv$ pc or $104\,\hinv$ pc at $z=2$).

Our six selected halos produce seven massive central galaxies at
$z=2$ that have no low-resolution particles within their virial
radius.  The galaxies span halo masses of $2\times 10^{11}-10^{12}
M_\odot$, stellar masses of $3.6\times 10^9-6.6\times 10^{10}M_\odot$,
and star formation rates (SFRs) from $5-60 M_\odot/$yr.  We label
them in order of increasing halo mass, and their global properties
at $z=2$ are listed in Table~\ref{table:sims}.
Note that given the stellar masses of our sample at $z=2$, it is likely that each of these galaxies
 will end up as an elliptical galaxy more massive than the Milky Way by $z=0$ \citep{Lu14,Lu15}.

\begin{deluxetable}{cccccc}
\tiny
\tablecolumns{6}
\tablewidth{0pc}
\tablecaption{Simulated Galaxy Sample at $z=2$ \, \, \, \, \, \, \, \, \, \, \, \, \, \, \, \, \, \, \, \, \, \, \, \, \, \, \, \, \, \, \, \, \, \, \, \, \, \, \, \, \, \, \,  \, \, \, \, \, \, \, \, \, \, }
\tablehead{\colhead{{\small Name}} &
	\colhead{{\small Symbol}} &
	\colhead{{\small $\Mhalo$}} &
	\colhead{{\small $\Mstar$}} & 
	\colhead{{\small SFR}} &
	\colhead{{\small $f_{gas}$}} 
}
\startdata
G1 & \includegraphics[width=0.02\textwidth]{figures/galaxySymbols/square.pdf}         & $1.91\times10^{11}$ & $3.57\times10^{9}$  & 4.93   & 0.54 \\
G2 & \includegraphics[width=0.02\textwidth]{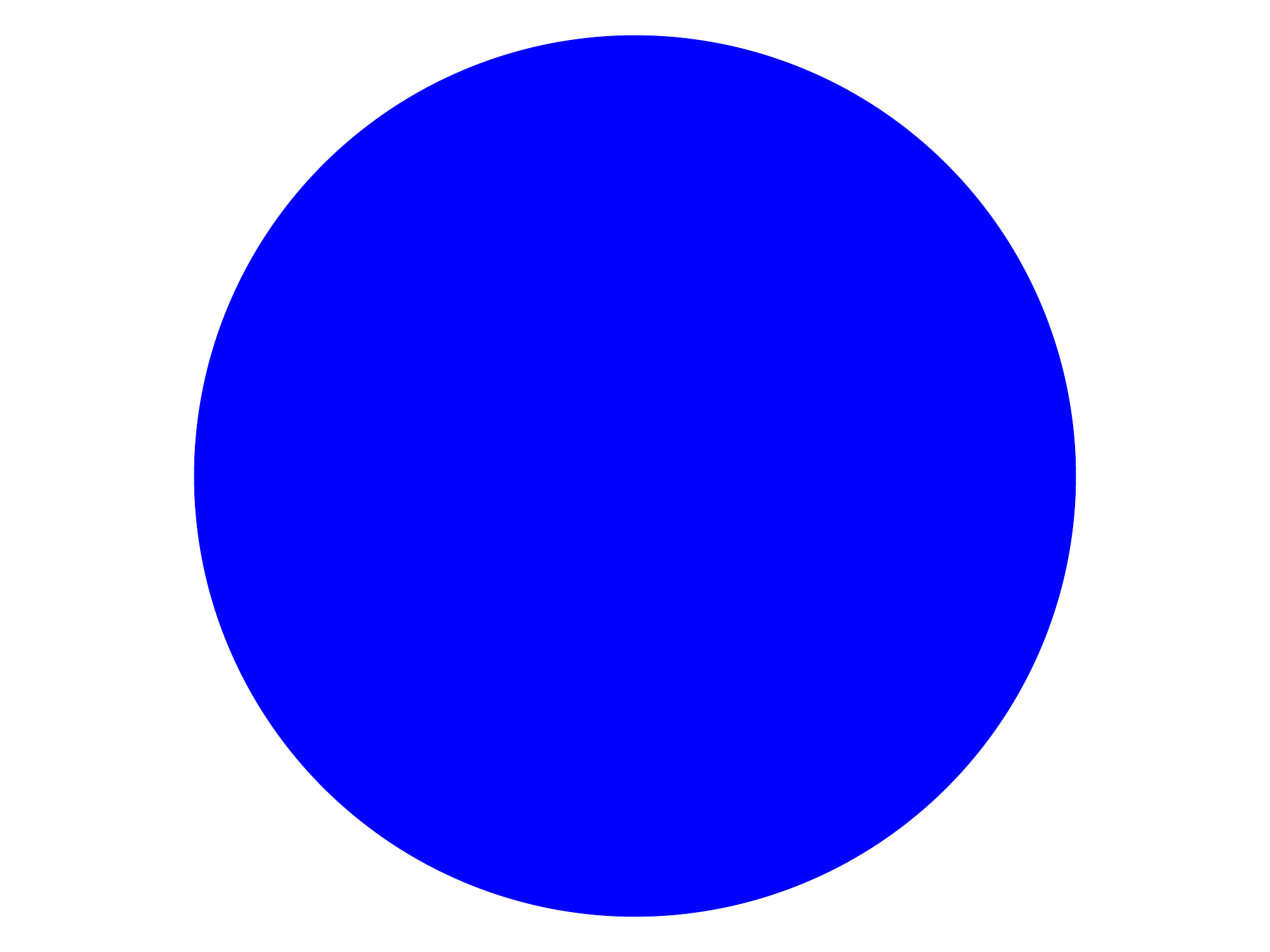}            & $1.98\times10^{11}$ & $7.77\times10^{9}$  & 9.96   & 0.47 \\
G3 & \includegraphics[width=0.02\textwidth]{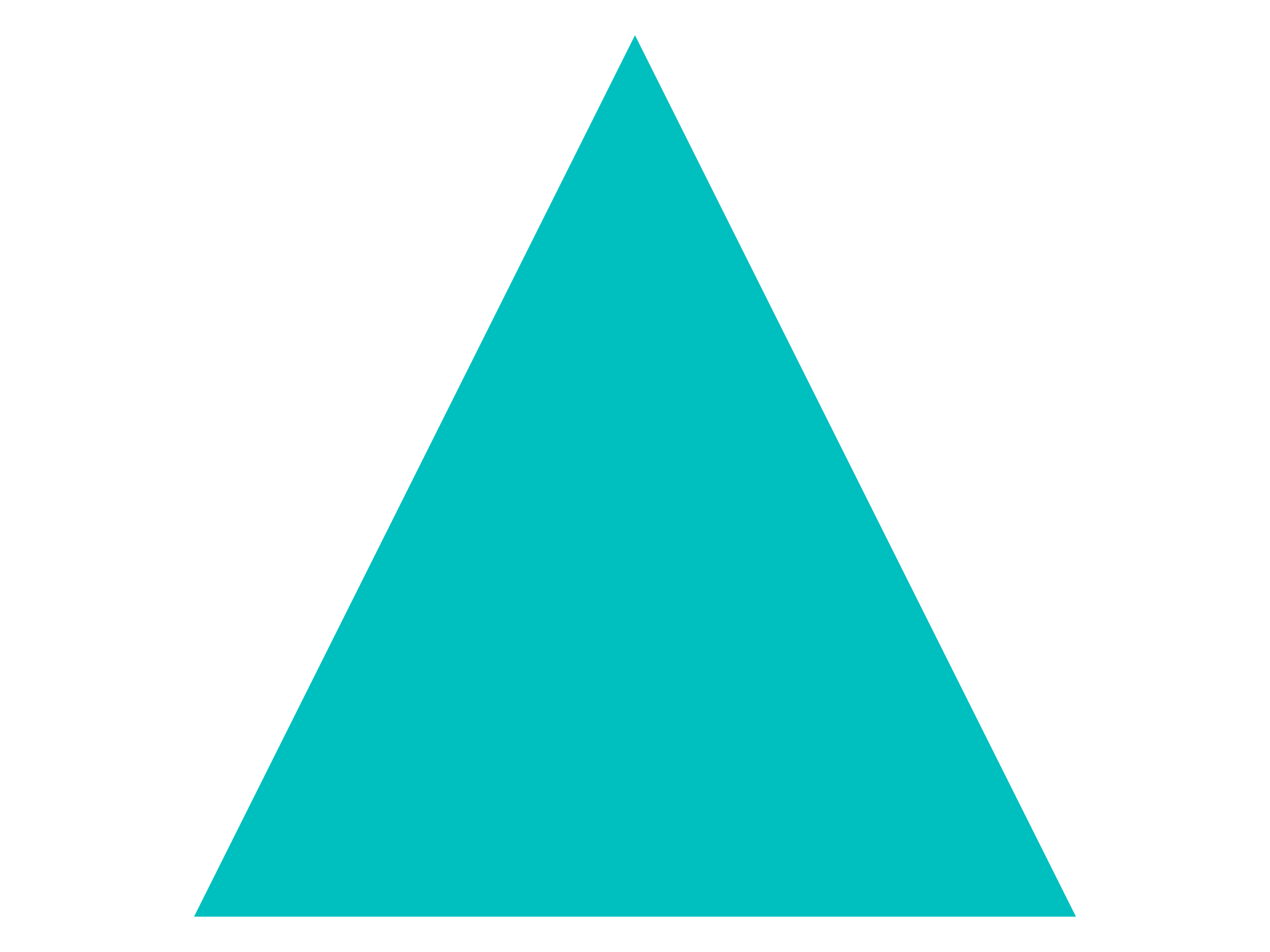}     & $3.35\times10^{11}$ & $9.49\times10^{9}$  & 8.77   & 0.57 \\
G4 & \includegraphics[width=0.02\textwidth]{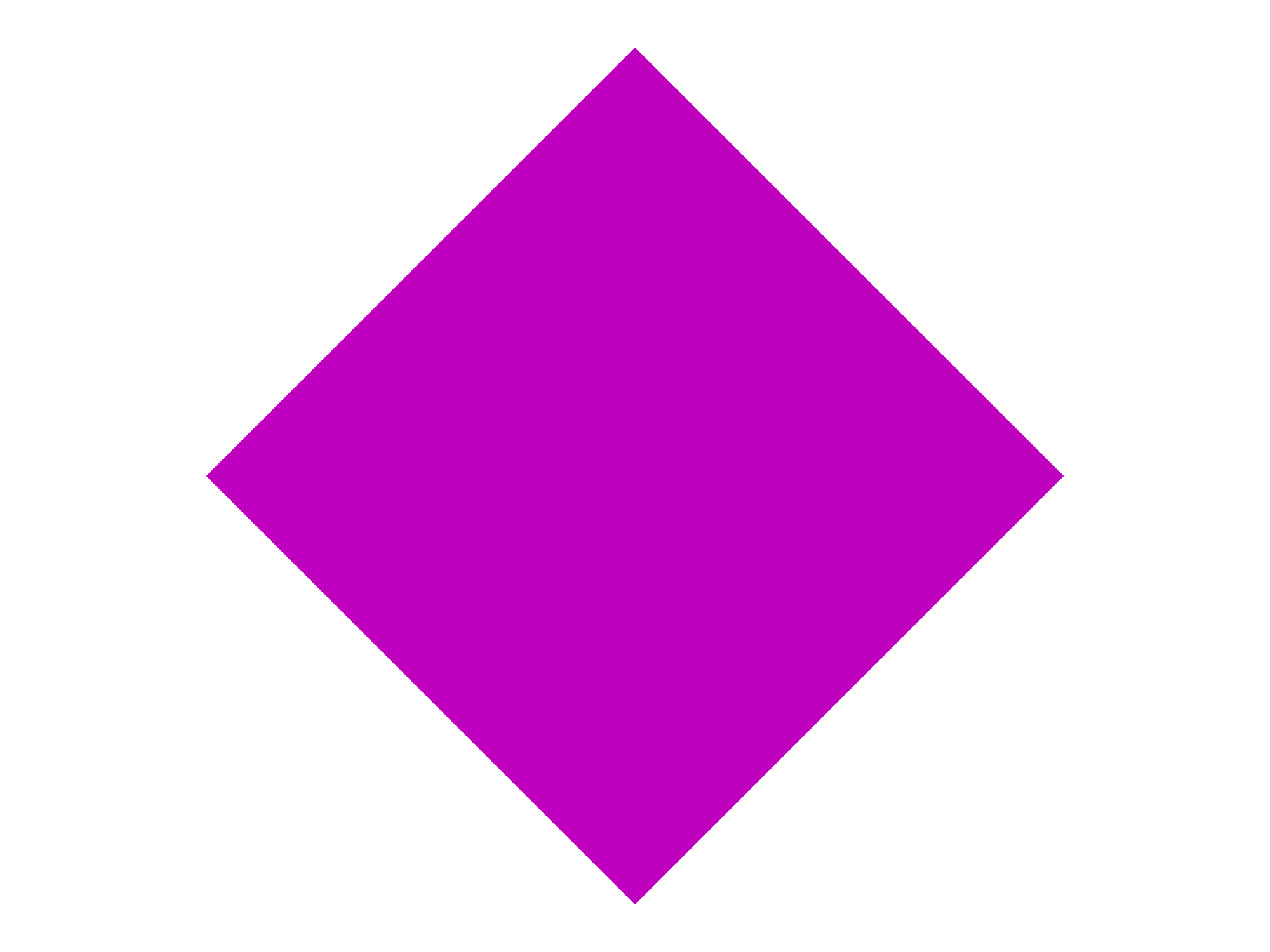}       & $4.85\times10^{11}$ & $1.80\times10^{10}$ & 2.51  & 0.44 \\
G5 & \includegraphics[width=0.02\textwidth]{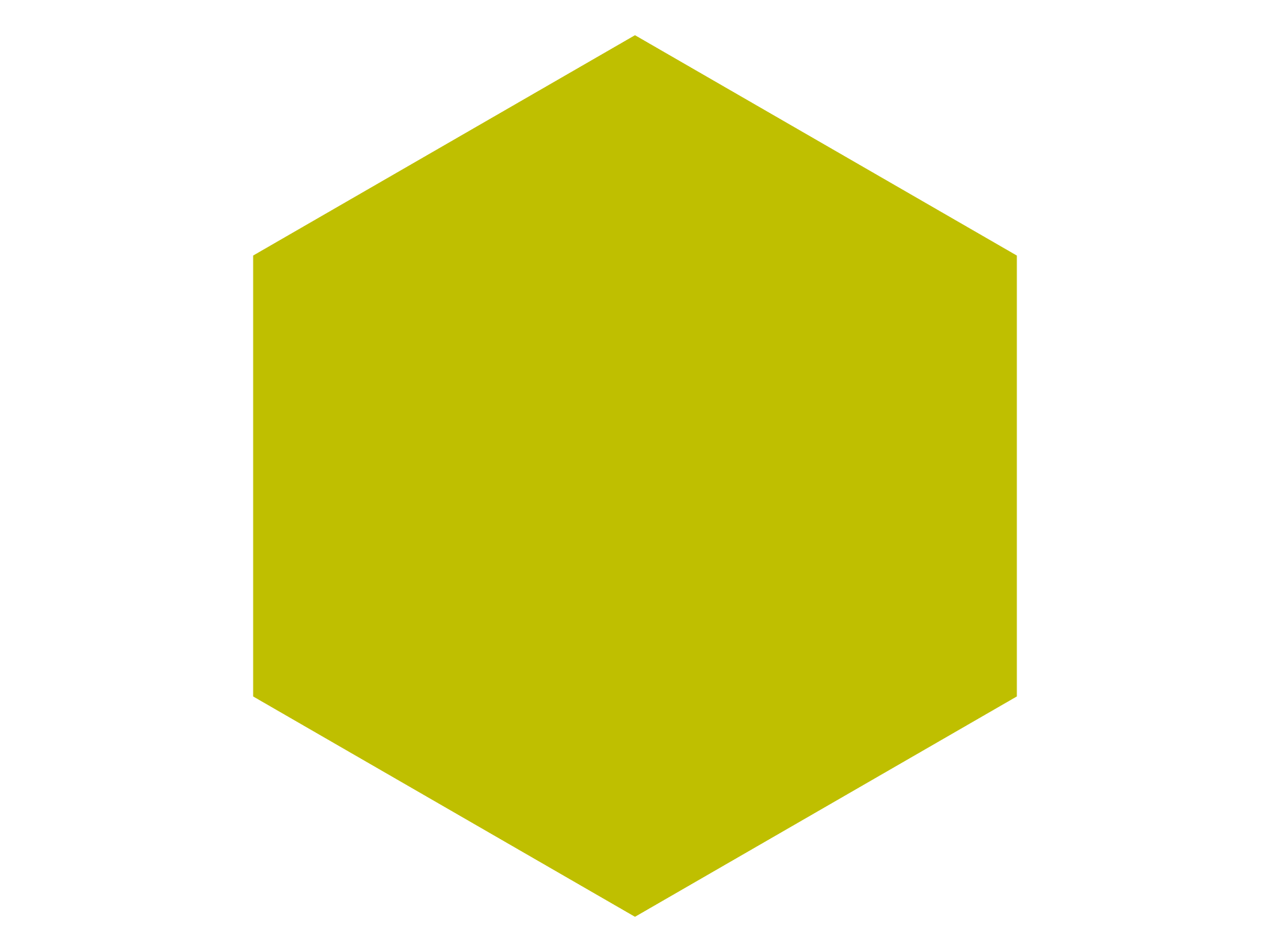}       & $9.55\times10^{11}$ & $4.05\times10^{10}$ & 19.93 & 0.39 \\
G6 & \includegraphics[width=0.02\textwidth]{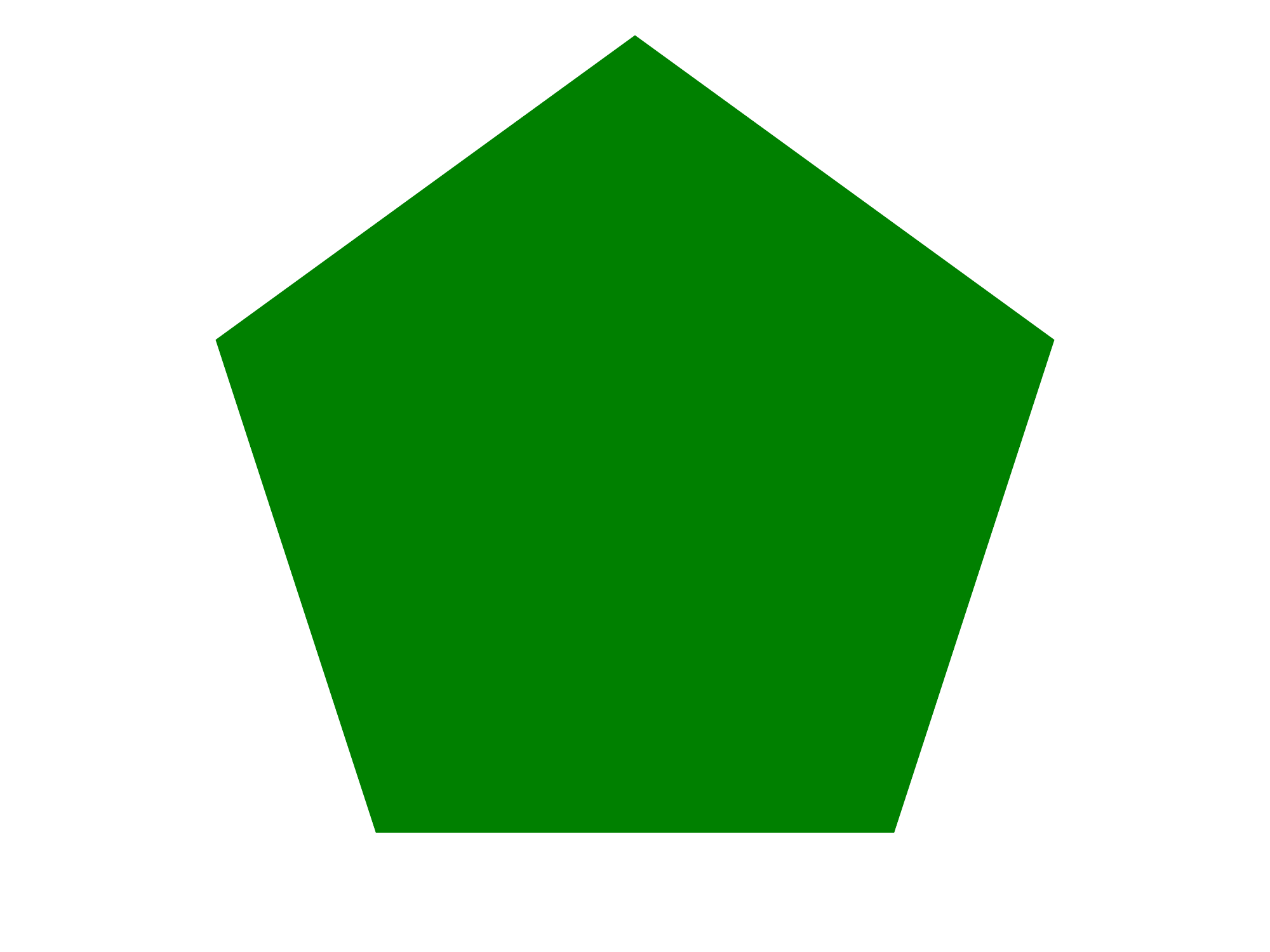}      & $1.15\times10^{12}$ & $5.58\times10^{10}$ & 37.50 & 0.21 \\
G7 & \includegraphics[width=0.02\textwidth]{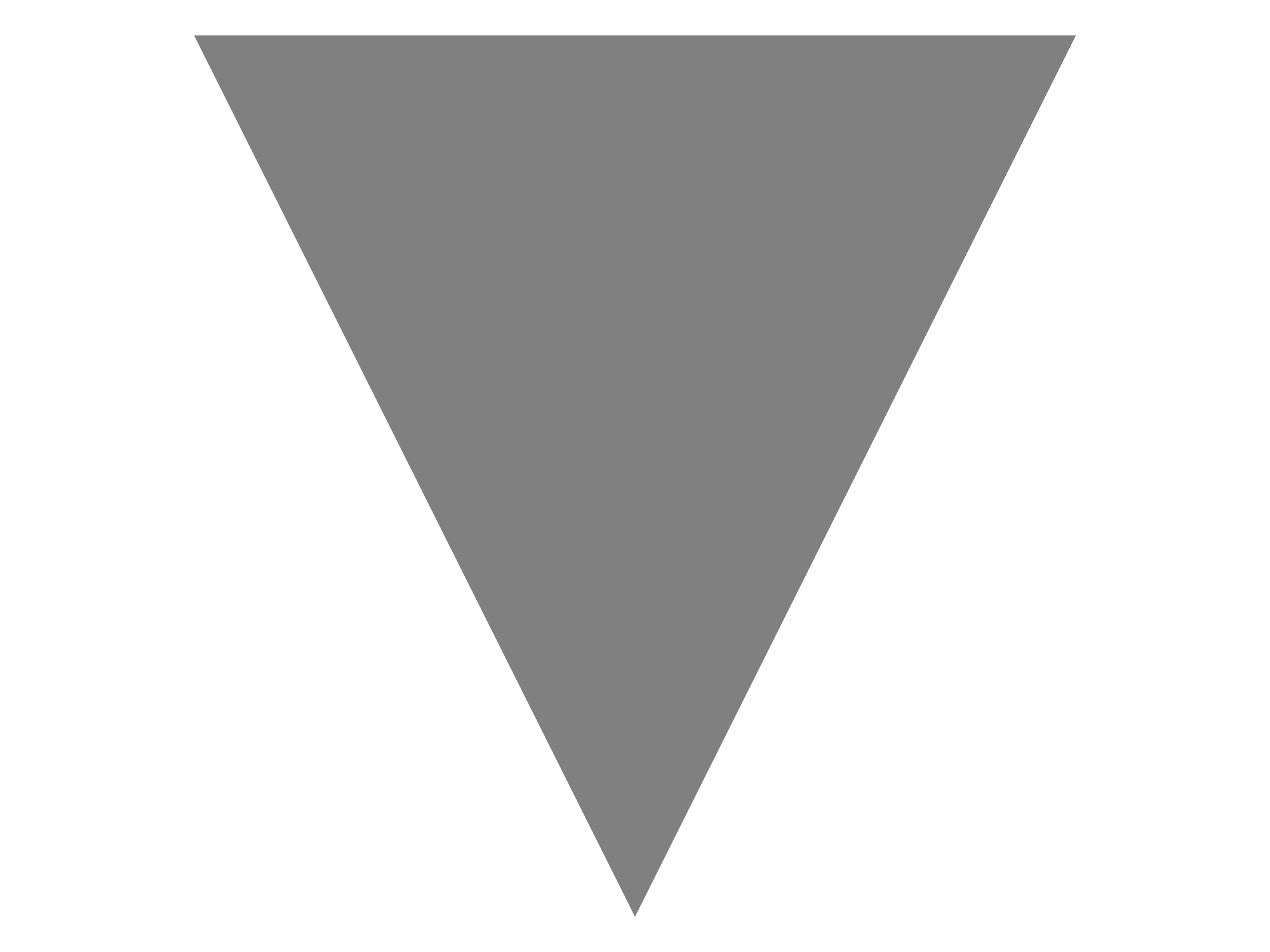} & $1.03\times10^{12}$ & $6.64\times10^{10}$ & 59.04 & 0.28 \\
\enddata
\label{table:sims}
\tablecomments{
Halos used throughout this work along with their central galaxies.
All were selected from a low resolution simulation in a
(16$\hinv \rm{Mpc})^3$ volume.
The smallest two halos (G1 and G2) have a comoving spatial resolution of 
$\epsilon_{grav}=313\hinv$ pc ($208\,\hinv$ pc at $z=2$), while the rest
(G3-G7) have $\epsilon_{grav}=625\hinv$ pc ($104\,\hinv$ pc at $z=2$).
$\Mhalo$ and $\Mstar$ are given in units of $\Msun$, 
while SFRs are given in units of $\Msun \,{\rm yr}^{-1}$.
}
\end{deluxetable}


\subsection{Simulation analysis with {\sc SPHGR}}
\label{sec:sphgr}

We perform the simulation analysis using the recently developed open-source
python package \textbf{S}moothed \textbf{P}article \textbf{H}ydrodynamics
\textbf{G}alaxy \textbf{R}eduction, or {\sc
SPHGR\footnote{\url{https://bitbucket.org/rthompson/sphgr}}}
\citep{SPHGR}.  In its most basic form, the code is responsible for
running a baryonic group finder to identify galaxies, a halo finder
to identify dark matter halos, assign galaxies to their
respective halos, calculate halo and galaxy global properties, and
finally iterate through previous time steps to identify the
most-massive progenitors of each halo and galaxy.  Information about
each individual halo and galaxy is collated in a convenient,
intuitive, and easy-to-access manner that eliminates unnecessary
overhead in the analysis process.

We use
\small{SKID\footnote{\url{http://www-hpcc.astro.washington.edu/tools/skid.html}}}
(Spline Kernel Interpolative Denmax) to identify galaxies as bound
groups of baryonic particles \citep{Governato97,SKID}.  All of the
galaxies presented in this work are well above our nominal 64 star
particle resolution limit \citep{Finlator06}.  To identify dark
matter halos and the associated baryonic particles we use the six
dimensional phase space halo finder {\sc
Rockstar-Galaxies\footnote{\url{https://bitbucket.org/pbehroozi/rockstar-galaxies}}}
\citep{ROCKSTAR}.  By considering particle velocities, {\sc Rockstar}
can much more accurately identify distinct groups of particles in
mergers when compared to conventional FOF methods
\citep[e.g.][]{Thompson15Bullet}.

After the identification of galaxies and halos via the above methods,
\sphgr creates individual python objects for each halo and galaxy.
These objects contain basic statistical information such as mass
and radius, and additionally the member lists of gas, star, and dark
matter (DM) particles (for halos) that belong to each object.  For
each {\sc SKID} galaxy, \sphgr assigns halo membership by
cross-correlating galaxy membership information with halo membership
to determine which halo has the most number of particles contributing
to that galaxy.  The most massive galaxy within each halo is
classified as the `central', while all others are classified as
`satellites'.  We calculate the H{\sc i} content of each galaxy
using the auto-shielding approximation as detailed in
\citet{Dave13}, along with global properties such as the stellar
mass, H$_2$ gas mass, star formation rate (SFR), and SFR-weighted
metallicity.

\sphgr identifies the primary progenitors of each galaxy and halo
throughout previous time steps, as follows.  For each halo we
identify the halo in the previous output that contains the most
dark matter particles from the current halo as the main progenitor.
In the case of galaxies, we utilise star particles rather than dark
matter particles to track progenitors in a similar manner.


\subsection{\sc Loser}
\label{sec:loser}

To create mock observations of our galaxies, we employ our newly
developed code {\sc Loser} (Line Of Sight Extinction by Ray-Tracing)
to compute the emission spectrum and magnitude maps, i.e. images,
for each galaxy.

{\sc Loser} uses the Flexible Stellar Population Synthesis library
\citep[FSPS;][]{Conroy09,Conroy10} to generate single stellar
population spectra interpolated to each simulated star particle's
age and metallicity.  Each star's spectrum is individually extincted
based on the integrated dust column by computing a line-of-sight
metal column density to the star towards a user-specified direction,
using an SPH kernel-weighted integral of particles along the LOS.
The metal column is converted to $A_V$ using relations measured for
the Milky Way \citep{Watson11} at solar metallicity and above, while
a metallicity-dependent dust-to-metal ratio is used at lower
metallicities based on high-z GRB measurements \citep{DeCia13}.
Given $A_V$, an extinction law is then applied; here we use a
\citet{Calzetti00} law.  {\sc Loser} also accounts for IGM attenuation
based on \citet{Madau95}, and nebular emission lines, though the former
is unimportant for the bands we consider, and the latter does not
play a significant role at the redshifts we consider.

The attenuated stellar spectra are then grouped into pixels of a
user-specified size.  User-selected band passes are then applied to
obtain rest- or observed-frame magnitudes in each pixel.  In our
case, we select our pixel size to correspond to $0.03"$ at the
appropriate redshift of the simulation output, which is comparable
to the pixel scale in drizzled~\citep{Koekemoer03} Wide Field Camera
3 (WFC3) images from {\it Hubble}.  The image is then blurred
according to the diffraction limit of the incoming wavelength; e.g.
for the case of F160W on WFC3 we apply a Gaussian smoothing to the
image with a FWHM of $\sim0.05"$. 

{\sc Loser} accurately estimates the emergent flux in bands where
reprocessed dust emission does not dominate, e.g. all the
rest-ultraviolet to near-infrared bands that we consider here.  {\sc
Loser} does not predict a spectrum for the far-infrared reprocessed
emission (though it does output the bolometric reprocessed luminosity)
as would be predicted using a full dust radiative transfer code such as
{\sc Sunrise}~\citep{Jonsson06}, but it is much faster and provides
greater user control over its input quantities such as the extinction
law.


\section{Global Properties}
\label{sec:globals}

To characterise our galaxy sample and ensure that they are broadly
in accord with observations, we first examine the evolution of their
global properties between $4\geq z \geq2$.


\subsection{Mass histories}
\label{sec:progen}

We begin by examining the mass growth histories of our galaxies,
and describe how we identify major mergers.  First, galaxies are
linked at $z=2$ with their most massive progenitor in previous
outputs as described in \S\ref{sec:sphgr}.  This allows for the
extraction of each galaxy's baryonic mass history, and hence we can identify
at each output (redshift) a merger ratio ${\cal R}$ defined by
\begin{equation}
{\cal R} \equiv \frac{M_1}{M_2} = \frac{M_t - M_{(t-1)}}{M_t},
\end{equation}
where $M_t$ is the mass of the galaxy at the current output, and
$M_{(t-1)}$ is the galaxy's mass in the previous output (outputs have an 
approximate spacing of $\Delta t \approx50-100$ Myr).
We consider merger ratios of at least 1:4 (${\cal R}\geq0.25$) 
to be major mergers.  We will not explicitly discuss
minor mergers, but canonically this is specified by ${\cal R}\geq 0.1$.
\
A histogram of ${\cal R}$ values within our sample separates into two components, one that peaks at ${\cal R}\sim 0.03$ and another that peaks at ${\cal R}\sim 0.4$, with a dip around ${\cal R}\sim 0.25$.  Hence our major merger cut at ${\cal R}\geq 0.25$ is chosen to isolate this separate population of events that is distinct from just an extension of typical ``smooth" accretion and minor mergers.

\begin{figure}
\includegraphics[width=0.5\textwidth]{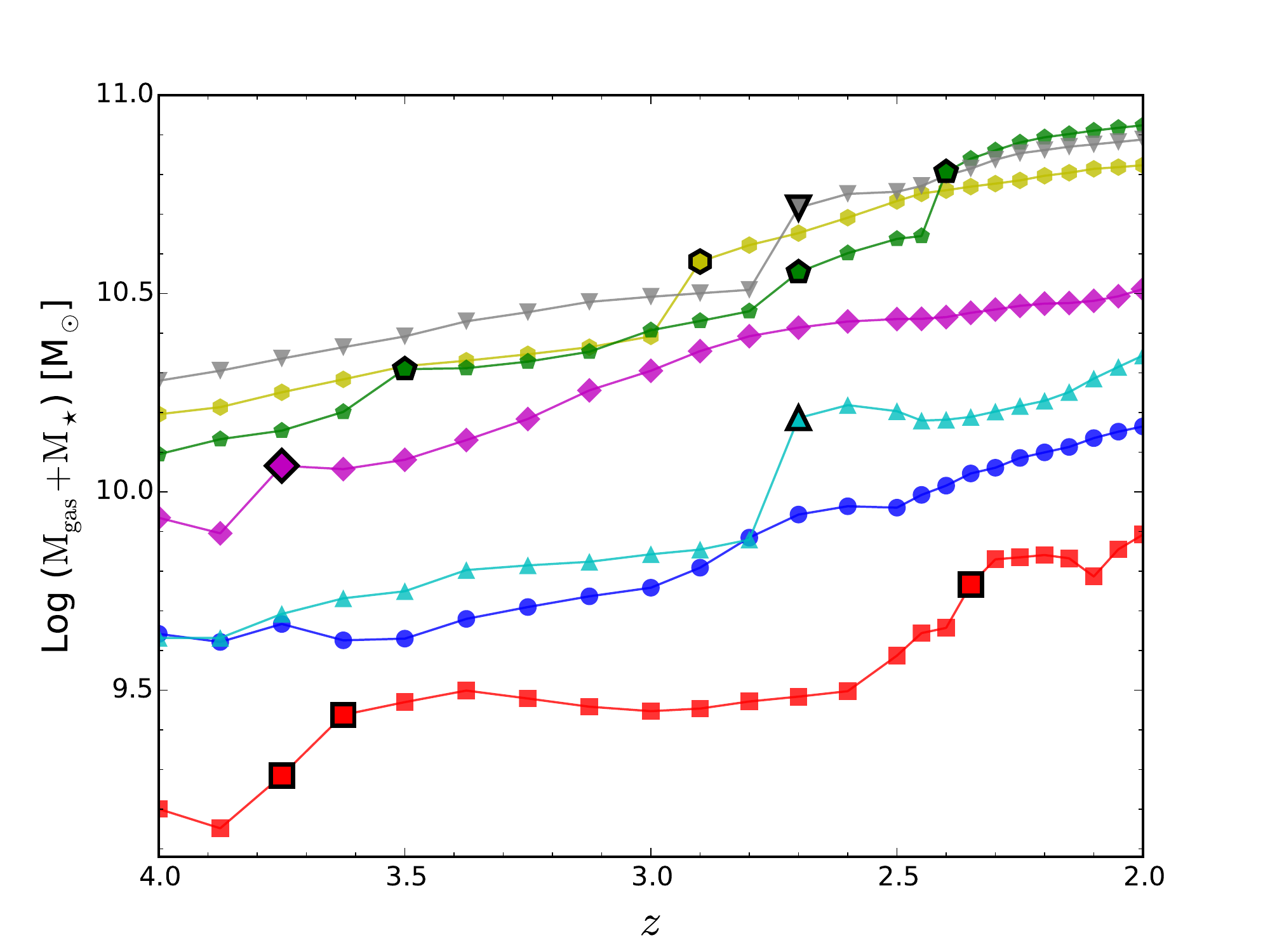}
\caption{
Evolution of the total baryonic mass ($M_{gas}+M_\star$) of each galaxy within
our sample between redshifts $4\geq z \geq2$.
The end of major mergers ($M_1/M_2\geq1/4$) are indicated by data points with black borders.
}
\label{fig:progen}
\end{figure}

Figure~\ref{fig:progen} displays the total baryonic mass
($M_{gas}+M_\star$) history of each galaxy within our sample.  The
ending time of identified major mergers are indicated by large
symbols with black borders.  The overall mass history of these
galaxies is dominated by relatively smooth growth~\citep{Murali02,Keres05},
but major mergers occur in all galaxies but one (G2) over this
redshift range.  There are a total of 10 major mergers that occur
within our sample between $4\geq z \geq 2$, yielding a typical
timescale between mergers of $\sim1.25\,{\rm Gyr}$.

\subsection{Gas fraction}
\label{sec:fgas}

\begin{figure}
\includegraphics[width=0.5\textwidth]{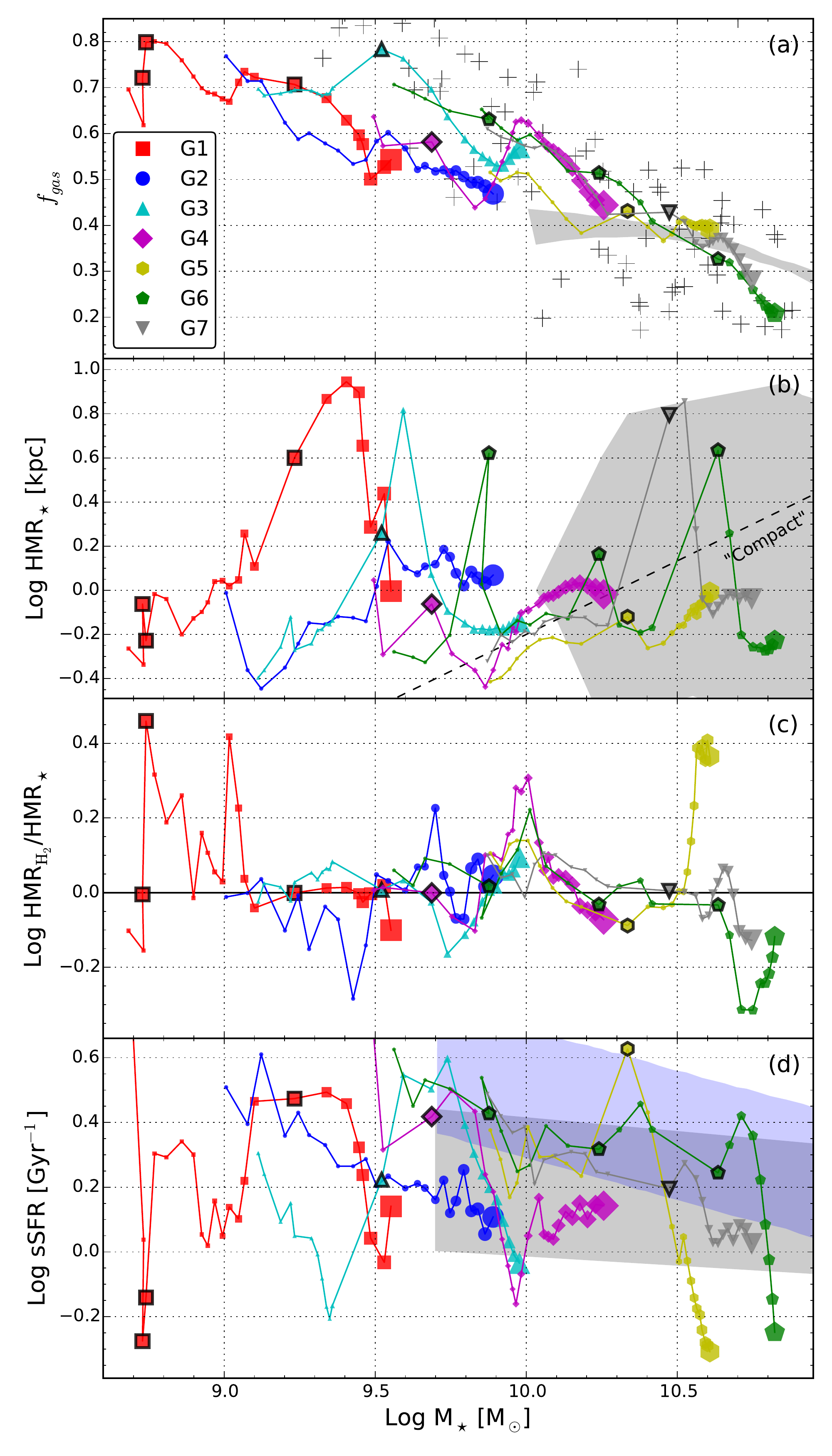}
\caption{
Redshift evolution of global properties of our galaxy sample between $4\geq z \geq 2$.
Major mergers (${\cal R}\geq1/4$) are again indicated by data points with black borders.
Panel~(a) shows $f_{gas}$ for each galaxy, with observational data at $z=2$ as the grey crosses \citep{Erb06}, and $z=1-1.5$ data from PHIBBS \citep{Tacconi13} as the grey contour.
Panel~(b) shows the half-mass radius for each galaxy, the `compactness' line for $1.4<z<3$ \citep{Barro13}, and $1<z<3$ observational data compiled by \citet{Bruce12} as the grey contour.
Panel~(c) shows the ratio of the molecular gas half-mass radius and the
stellar half mass radius for each galaxy, the `compactness' line for $1.4<z<3$ \citep{Barro13}, and $1<z<3$ observational data compiled by \citet{Bruce12} as the grey contour.
Panel~(d) shows the specific star formation rate for each galaxy with the grey contour representing observational data at $z=2$ \citep{Daddi07,KrumholzDekel12}, and the blue contour  from \citet{Speagle14}.
Panel~(d) shows the specific star formation rate for each galaxy with the grey contour representing observational data at $z=2$ \citep{Daddi07,KrumholzDekel12}, and the blue contour  from \citet{Speagle14}.
}
\label{fig:globals}
\end{figure}

The gas fraction ($f_{gas}\equiv M_{gas}/(M_{gas}+M_\star)$) history
of our galaxies is shown in Figure~\ref{fig:globals}(a), again with
the end of major mergers shown as points with black borders for
reference.  Overall, the gas fraction evolves downwards with time.
As discussed in \citet{Dave11b}, this owes to efficient
outflows in small galaxies at early epochs that keeps galaxies from
converting gas into stars.  After major mergers, we often see a somewhat
more rapid decrease in $f_{gas}$, suggesting that these galaxies
consume the gas available for star formation quickly during these events.

The predicted values of $f_{gas}$ between $4\geq z \geq2$ for our
galaxies broadly agree with the \citet{Erb06} $z=2$ observations
shown as grey crosses, with galaxies evolving mostly down along the
relation with time.  This indicates that the ISM gas content of our
galaxies is generally in accord with observed galaxies at these
epochs.  It is, however, somewhat higher than the gas fractions in the
lower redshift $z\approx 1-1.5$ PHIBBS data \citep[shown as the
grey contour;][]{Tacconi13}. 

\subsection{Size}
\label{sec:size}

Figure~\ref{fig:globals}(b) shows the stellar half-mass radius
(HMR$_\star$) of each of our galaxies between $4\geq z \geq2$.
Galaxies typically have sizes around 1 physical kpc at all epochs,
and show no obvious trend with mass.  The weak evolution of galaxy
size with respect to stellar mass is in broad agreement with numerical
models and observations out to $z\sim 1$
\citep[e.g.][]{Somerville08,Brooks11}.  Our galaxy sizes also lie
within the locus of $3 > z > 1$ observations shown as the grey
contour \citep{Bruce12}.  We note that our physical resolution (i.e.
gravitational softening length) at these epochs is generally better
than $\sim 200$~pc, so these galaxies are all reasonably well-resolved.

We also overlay the $3>z>1.4$ line that delineates ``compact"
galaxies from \citet{Barro13}.  Identifying the progenitors of
massive compact galaxies has been a topic of great interest recently
\citep[e.g.][]{Barro13,Patel13,Stefanon13,Wellons15,Zolotov15,Fang15,Stringer15,Graham15}.
In our simulations, galaxies grow much more slowly in half-mass
radius than in stellar mass; in this plot they tend to move nearly
horizontally with time.  As a result, the three most massive galaxies
(G7, G6, G5) all end up in the compact regime.  G4 undergoes an
early merger that lowers its radius into the compact regime at a
low mass, but it quickly re-grows out of being compact, then nearly
returns to becoming compact by $z=2$.

Mergers often greatly but temporarily increase the size.  This may
be because SKID will often identify two merging galaxies as a single
one prior to final coalescence, resulting in a large half-mass
extent.  However, by the next output the galaxy usually settles
back to something similar to its previous size.  In detail, the final
radius tends to be slightly smaller than the pre-merger one (though
not always).  Nonetheless, it is not obviously the case that major
mergers are primarily responsible for driving galaxies into the
compact regime.  Instead, the primary driver seems to be simply the
fact that galaxy growth in stellar mass is not accompanied by a
sizeable increase in radius at these epochs, which thus eventually
drives galaxies into the compact regime.

A separate question is what process removes the star-forming gas
once it reaches the compact regime to produce red and dead
compact ellipticals as observed~\citep{vanDokkum10}.  We cannot
address this question directly in our simulations since we do not
include a model for quenching star formation.  Nonetheless, if such
quenching is associated with the presence of massive ($M_h\ga
10^{12}M_\odot$) halos filled with hot gas as is often
speculated~\citep[e.g.][]{Croton06,Gabor14}, then it may be that
galaxies coincidentally achieve sufficiently massive halos roughly
at approximately the same stellar mass where their radii move into
the compact regime.  In such a scenario, most quenched galaxies
would start out as compact, although the exact fraction depends on
the stochasticity in the relationship between quenching and halo
mass (or whatever the governing parameter might be).  We leave a more
detailed study of this for future work.

\subsection{Gas radius}
\label{sec:rgas}

Figure~\ref{fig:globals}(c) shows the ratio of the molecular (star-forming) gas
half-mass radius and the stellar half-mass radius.  Overall, the
two evolve together showing, as expected, that the stars form where
the gas is located.  However, there are some interesting departures.  Shortly
after a merger, G4's molecular half-mass radius increases to twice
its stellar half-mass radius, before settling back down in size.  This
mostly owes to the stellar radius dropping just prior, suggesting
a scenario where internal dynamical effects condense the core before
gas falls back in to regrow a larger disk, which then forms stars
allowing the stellar radius to catch up.  Such effects have been
seen in isolated merger simulations where the progenitors are
gas-rich~\citep{Robertson04} as well as cosmological zoom simulations
of early galaxy growth~\citep{Governato07}.  

Another interesting case is G5, where just before $z=2$ its gas
radius shoots up to more than twice its stellar radius.  Again,
this happens not too long after a merger when the stellar radius
drops a bit, suggesting a similar scenario of ongoing disk regrowth.
This suggests that if the molecular gas radius can be measured,
e.g. using ALMA, and compared to the stellar radius from {\it HST},
outliers in this relation may identify galaxies that have recently
undergone internal dynamical evolution via a merger, and are in the
phase of disk re-growth.


\subsection{Specific star formation rate}
\label{sec:ssfr}

In Figure~\ref{fig:globals}(d) we plot the specific star formation
rate (sSFR~$\equiv~SFR/M_\star$) of our sample, with the grey contour
representing $z\approx 2$ observational data from \citet{Daddi07},
and the blue contour representing a fit to observations by
\citet{Speagle14}.

Galaxies generally show a mild downward evolution in sSFR as a
function of $M_\star$.  For the vast majority of the time, all our
galaxies lie within the observed range, which evolves
slowly at $z\ga
2$~\citep[e.g.][]{KrumholzDekel12,Stark13,Duncan14,Speagle14}.  At
the final redshift, all the galaxies but G5 and G6 lie within the range
of the observations.  This is a non-trivial success that is difficult
to achieve in lower-resolution
simulations~\citep{Dave08,Sparre15,Somerville15}, and suggests that
the higher resolution, and potentially the inclusion of a molecular
hydrogen-based star formation prescription, is important for obtaining
sufficiently high sSFRs at $z\sim 2$.

For G5 and G6, these galaxies mostly evolved within the observed
range, but experienced fairly sudden drops in sSFR just prior to
$z=2$.  This was preceded by a mild increase in sSFR associated
with a merger event.  The drop in sSFR occurs along with a rise
in the ratio of the gas to stellar radius.  Hence for these objects
a plausible scenario is, as discussed earlier, that they underwent
a merger and consumed their dense gas, and are now regrowing a gas
disk that has not yet achieved a sufficient mass or density to restart
vigorous star formation.  These may be examples of the sort of
``quenching" discussed in \citet{Feldmann15}, but it is not quenching
in the conventional sense that results in red-and-dead galaxies, since
eventually the gas accretion will likely re-invigorate star
formation~\citep[e.g.][]{Gabor10}.

In summary, our galaxies lie within the range of typical observed
galaxies at $z\sim 2$ in terms of their gas content, size, and
specific SFR.  Hence they provide a plausible set of objects with
which to examine morphological characteristics and their correlation
with merger history.


\section{Morphological statistics}
\label{sec:morphstats}

We now examine non-parametric morphological statistics for our
galaxies at $4\geq z \geq 2$.  Every galaxy in Table~\ref{table:sims}
is processed through {\sc Loser} six times for each of our 24
outputs between $4\geq z \geq 2$; three correspond to random
orientations aligned with the axes of our simulation box, while the
other three are composed of a face-on orientation, and two edge-on
orientations separated by $90^\circ$.  These viewing angles will
be represented by the colours cyan (random), blue (face-on), and red
(edge-on), respectively.  This results in 144 mock observations for
each of our seven galaxies, or 1,008 total mock observations for
our entire sample.

We focus here on the F160W band as the reddest WFC3 band that
provides the most direct correlation with stellar mass.  We have
also repeated this analysis using the F814W band of ACS, which falls
in the rest-frame ultraviolet, but the statistics and correlations
we obtain are not substantially different and the resulting conclusions
remain the same; hence we do not discuss F814W further here.  We will
discuss detailed properties of colour variations in future work, and
confine ourselves here to the F160W images for calculating our
non-parametric morphological statistics.

\subsection{Basic morphological characteristics}

\begin{figure*}
\includegraphics[width=1.1\textwidth]{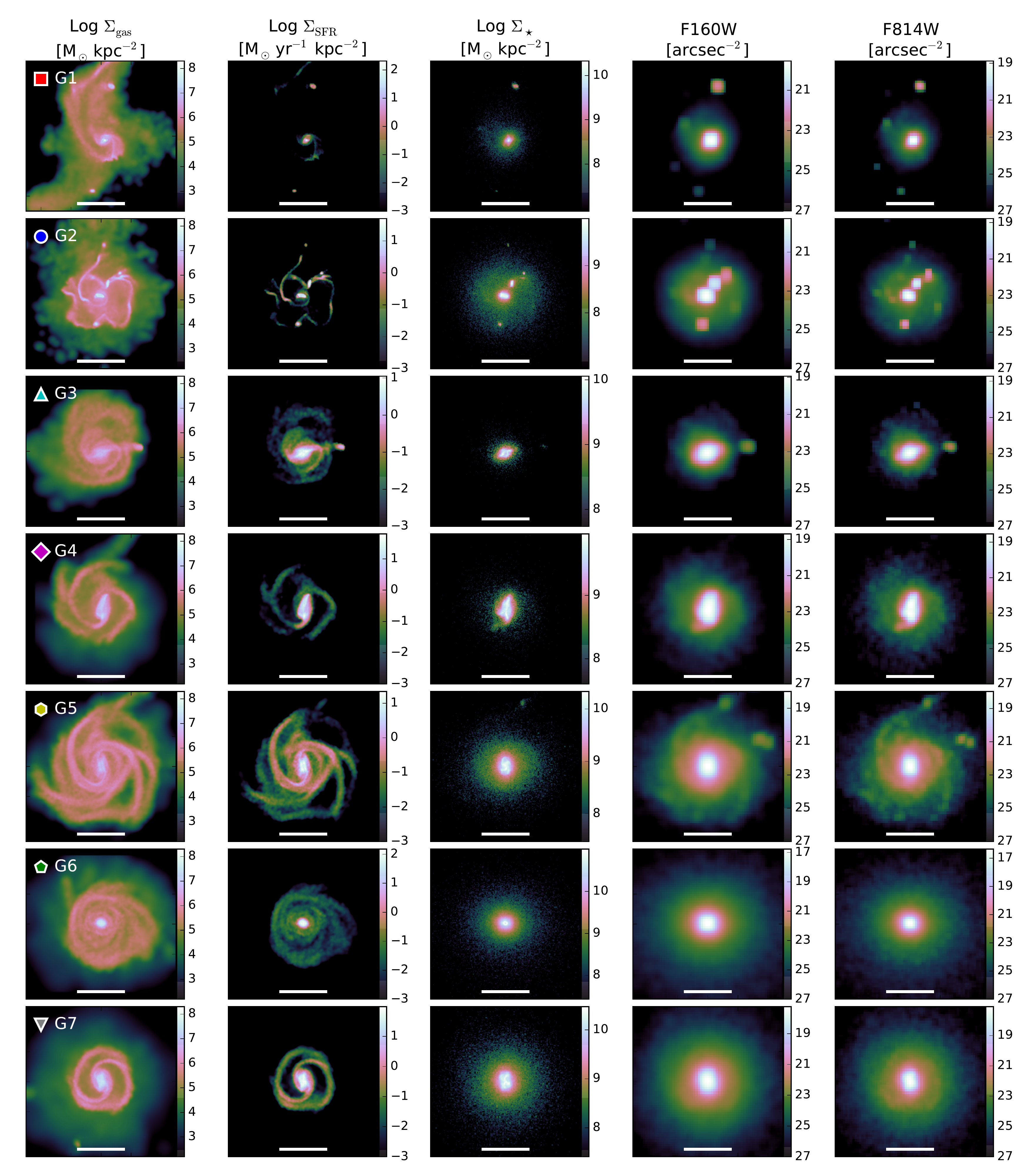}
\caption{
Face-on data for each galaxy from Table~\ref{table:sims} at $z=2$.
Columns 1-3 show surface density features, while columns 4 and 5 show each galaxy processed through {\small LOSER} in the F160W and F814W bands, respectively,
as described in \S\ref{sec:loser}.
The white scale bar corresponds to $5$ physical kpc.
}
\label{fig:postage}
\end{figure*}

To set the stage, Figure~\ref{fig:postage} shows face-on processed
images from {\sc Loser} for each galaxy in our sample at $z=2$.
The three columns show the gas, star formation rate, and stellar
surface densities, respectively, while the final two columns show
the extincted {\sc Loser} images in the WFC3 H-band (F160W) the ACS
I-band (F814W) filters.  The flux from images like these are used
to calculate their morphological statistics in the following sections.
The white bar at the bottom of each image represents a scale of 5
physical kpc.

Overall, the gas surface densities clearly indicate a disk-like
morphology, albeit with significant levels of disturbance evident
owing to ongoing minor mergers.  The SFR generally follows the
densest gas, mainly occurring in gas with $\Sigma_{\rm gas}>10^7
M_\odot$kpc$^{-2}$ in accord with the \citet{Kennicutt98} relation.
The stellar component does not show as obvious a disk-like
morphology, though we have checked that they do show rotation.  The
stellar map also shows evidence for ongoing merging activity, most
strongly in the smaller galaxies such as G2.  The gas disks are
quite extended compared to the stars, reaching 20~kpc across.

In the mock images, even at {\it HST}'s resolution, many
of the small-scale features in these galaxies are washed out.
Nonetheless, the fact that G2 is about to undergo a merger clearly remains
evident; interestingly, this is the only one of our galaxies that
does not show an identified merger from $4<z<2$.  Spiral arms are
barely if at all evident. They are most apparent in G5, which resembles the
grand design spiral seen by \citet{Law12}.

\subsection{Concentration and Asymmetry}
\label{sec:ca}

A basic morphological parameter of a galaxy is its light concentration
($C$). This has long been used to morphologically characterise
galaxies in various classification schemes \citep[e.g.][]{Morgan57}.
It has also been used extensively to quantify trends in local
galaxies \citep[e.g.][]{Bershady00,Conselice03}.  Higher values of
$C$ mean that a larger amount of the galaxy's light is contained
within the central region.  Concentration is typically defined as
the ratio of light within a circular radii containing 20\% ($r_{20}$)
and 80\% ($r_{80}$) of the total galaxy flux \citep{Conselice03}:
\begin{equation}
C=5 \log\left(\frac{r_{80}}{r_{20}}\right).
\end{equation}
The concentration is calculated using pixels within $1.5 r_p$ where $r_p$ is the
Petrosian radius, which we define as the radius of a circular aperture
where the flux at that radius is 20\% of the flux within that radius
\citep[e.g.,][]{Bershady00,Conselice03}.  

Asymmetry ($A$) is another common morphological statistic that
quantifies the rotational symmetry of a galaxy
\citep[][]{Abraham96,Brinchmann98,Conselice03}.  It is measured by
rotating the image by $180^\circ$ and subtracting that from the
original image \citep{Abraham96,Conselice00}:
\begin{equation}
A = \frac{\sum |f_0 - f_{180}|}{\sum |f_0|},
\label{eq:asymmetry}
\end{equation}
where $f_0$ is the original image pixel flux values, and $f_{180}$
is the rotated image's pixel flux, and the sum is over all pixels within
$1.5 r_p$.  Higher values of $A$ correspond to higher degrees of
asymmetry, which can be used to identify merger-candidates.
\citet{Conselice03,Conselice08}, for example, set a threshold value of
$A\geq0.35$ for major mergers based on their studies of nearby galaxies in
various phases of evolution.
They focused on numerous different data sets that included normal galaxies, 
starburst galaxies, dwarf irregulars, dwarf ellipticals, and galaxies
undergoing a merger.

\begin{figure}
\includegraphics[width=0.5\textwidth]{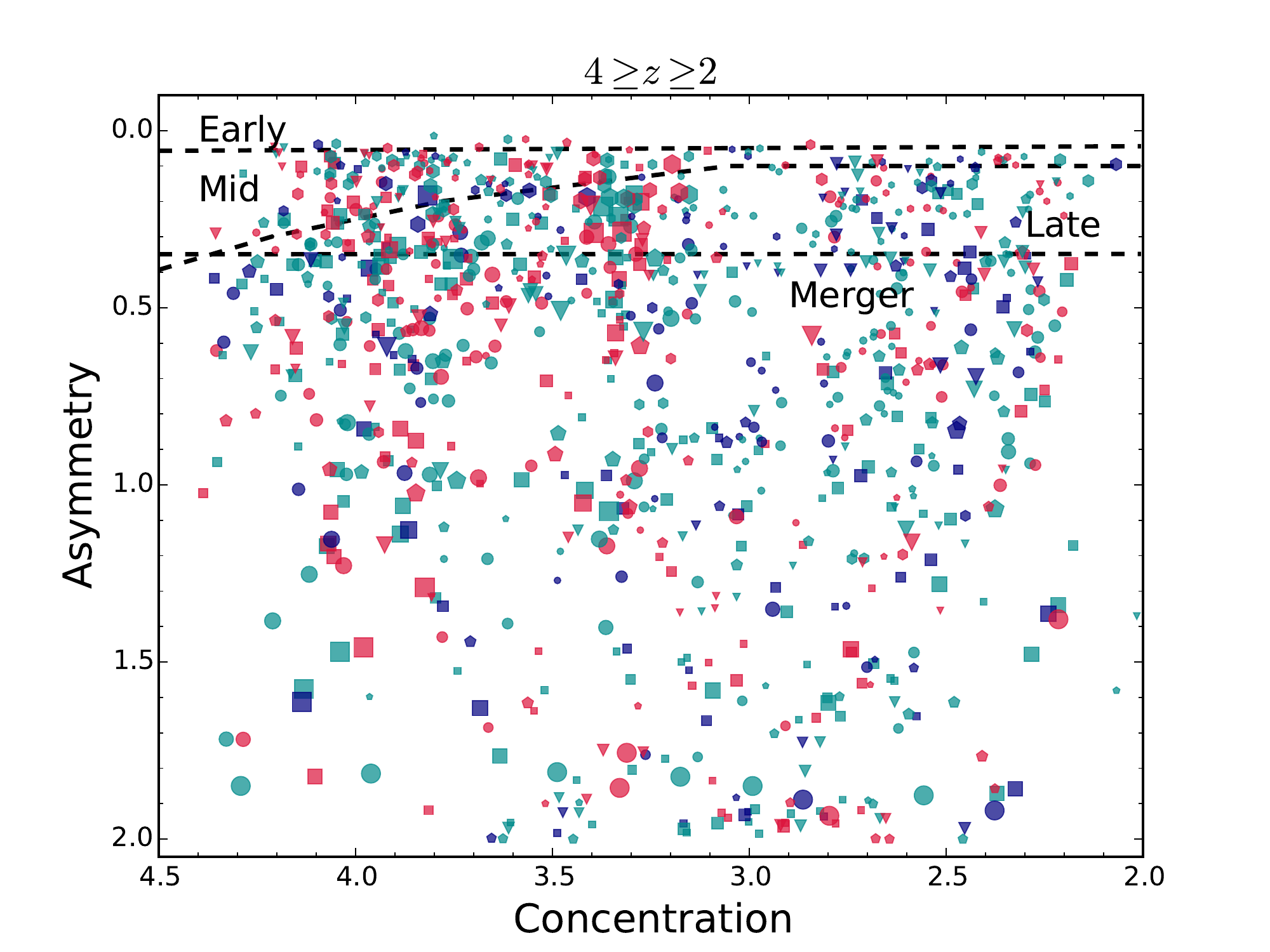}
\caption{
Asymmetry $A$ as a function of Concentration $C$ for our sample between $4\geq z \geq 2$.
Each galaxy in our sample contains 6 data points: 3 random viewing orientations (cyan), 2 edge-on views separated by $90^\circ$ (red), and one face-on view (blue).
Point size corresponds to the specific star formation rate (sSFR) of each galaxy, with smaller points representing smaller sSFRs.
Dashed lines correspond to regions in which different galaxy types are typically found in the local universe \citep{Bershady00,Conselice03}.
}
\label{fig:ca}
\end{figure}

$C$ and $A$ are traditionally plotted against each another to classify
a galaxy's morphology and merger state.  Figure~\ref{fig:ca} shows
$C$ vs. $A$ for our 1,008 mock observations between $4\geq z\geq
2$.  The lines indicate the region in which different galaxy types
are typically found in the local universe \citep{Bershady00,Conselice03},
and the point sizes are representative of each galaxy's sSFR.  

Within our sample we find a median concentration of 3.3, and a
median asymmetry of 0.49.  Hence using the canonical definition
that identifies mergers as objects with $A\geq0.35$, the median
galaxy at $z=2-4$ in our sample is a merger.  Given that major
mergers are not that frequent as shown in our mass history plots,
this indicates that the canonical value, tuned for lower-redshift
galaxies, is not straightforwardly applicable at higher
redshifts.  We will examine this more quantitatively later.

Examining each of these quantities individually allows us to
investigate potential correlations with the merger ratios extracted
from the simulation (\S\ref{sec:globals}).  We split our data
into three groups corresponding to the face-on, edge-on, and random
viewing angles.  Random orientation is the most directly relevant
when comparing to the full sample of objects in a galaxy survey,
but it may be possible to isolate a sample of face-on or edge-on
objects for comparison to the other panels.  Two power laws are fit
to the data within each panel; one with no weighting (blue), and
the other weighted by the galaxy's sSFR (orange).  The sSFR weighting
can potentially assist merger identification since mergers tend to
have higher sSFRs, and it is also information typically available
in a multi-wavelength galaxy survey.  Each panel's legend shows the
power-law index of each fit along with the uncertainty in the slope.

\begin{figure*}
\includegraphics[width=1\textwidth]{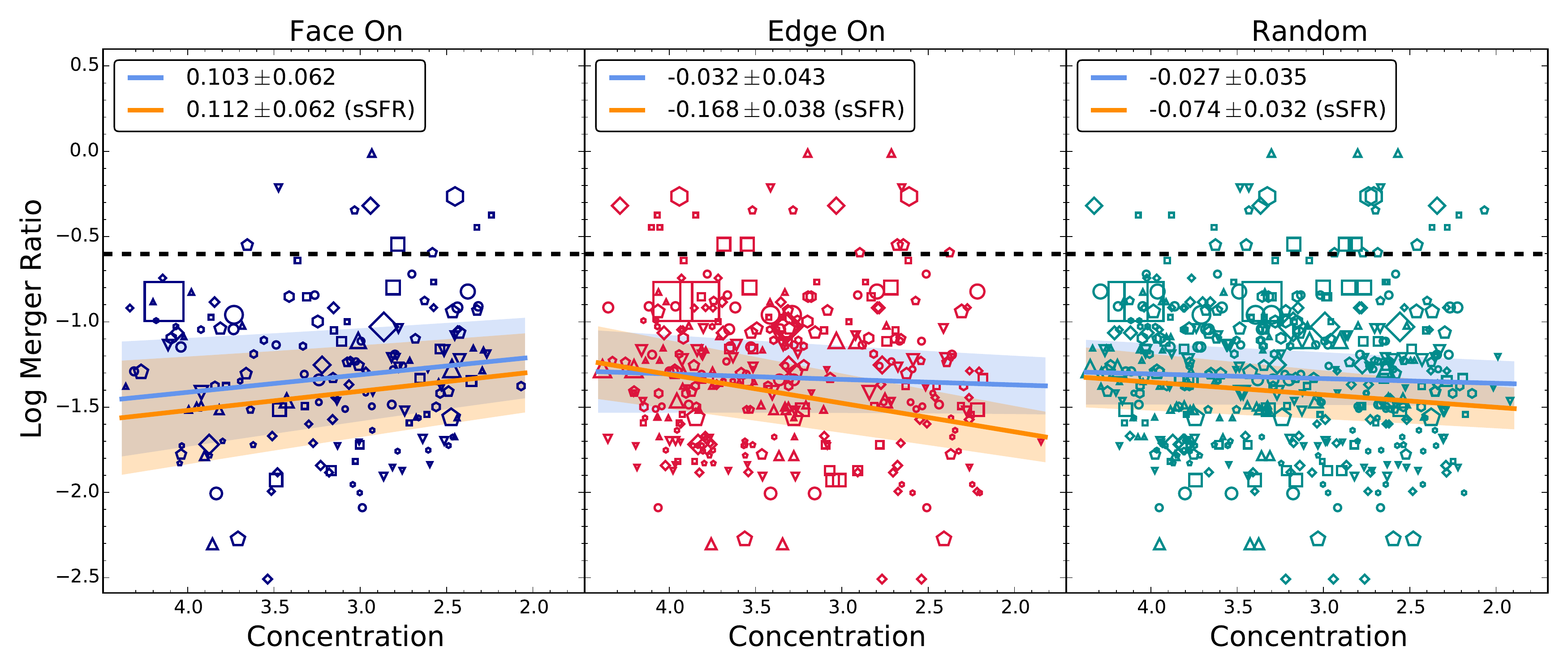}
\caption{
The merger-ratio $R$ as a function of the concentration ($C$) parameter 
discussed in \S\ref{sec:progen} split into different panels corresponding to the different viewing angles.  
Point shapes correspond to different galaxies (see Table~\ref{table:sims}) while point size corresponds to a galaxy's sSFR.
The dashed line represents the threshold for major mergers with a mass ratio of ${\cal R}=0.25$.
Two power laws are fit to each panel and shown with $1\sigma$ error contours; the blue is a result of fitting with no weighting, and the red fit is weighted by the sSFR (\S\ref{sec:ssfr}). 
Legends indicate the power law index with errors.
}
\label{fig:concentrationMR}
\end{figure*}

Figure~\ref{fig:concentrationMR} shows the merger ratio as a function of the 
concentration of each galaxy for each viewing orientation.
The horizontal dashed line shows the value of ${\cal R} = 0.25$ above which we
identify the galaxy as having undergone a major merger since the
immediately previous output.
From this it is clear that there is hardly any correlation between
the concentration parameter and our galaxy sample's merger state.
This is not entirely surprising, as the major merger cut criteria
is usually defined in terms of $A$ (Figure~\ref{fig:ca}).  The only
significant correlation occurs in the face-on sSFR-weighted fit,
but this shows an {\it anti-}correlation since one expects mergers
to increase the concentration. 

\begin{figure*}
\includegraphics[width=1\textwidth]{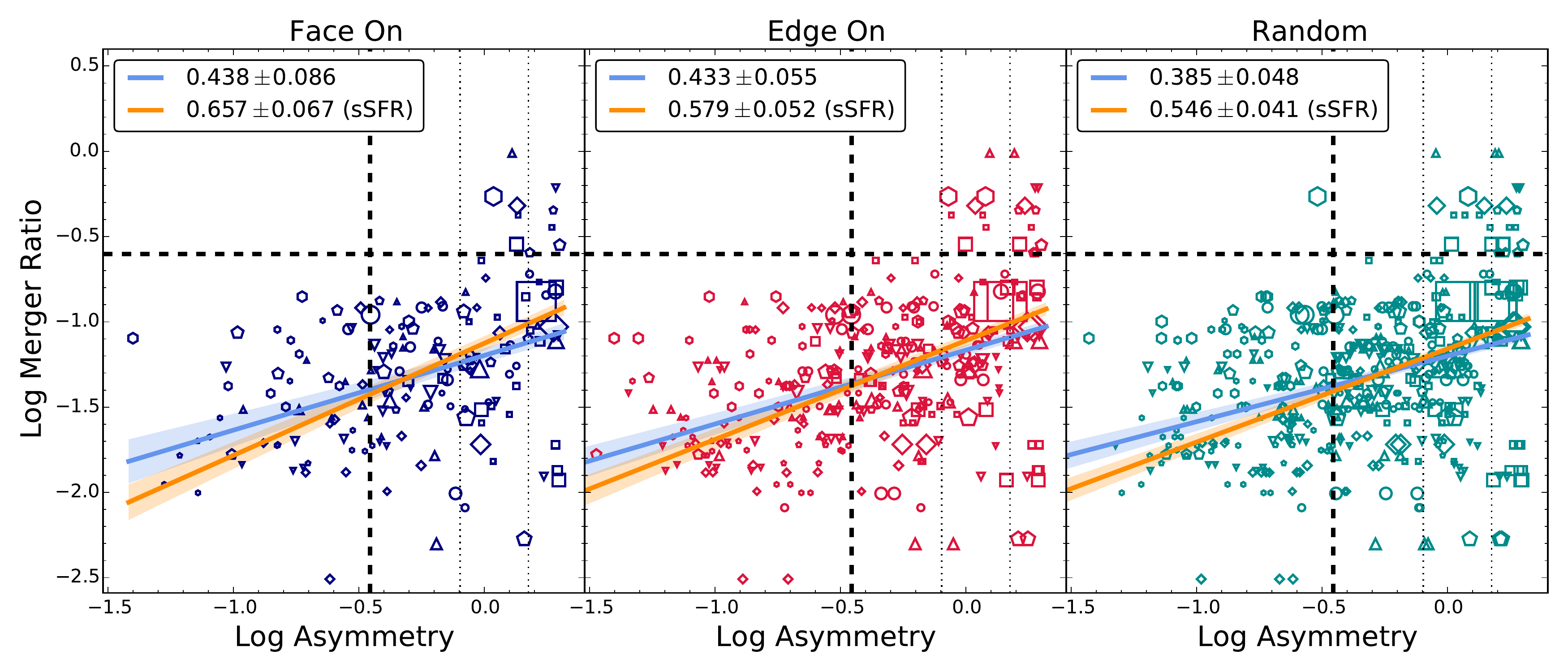}
\caption{
The merger-ratio $R$ as a function of the asymmetry ($A$),
similar to Figure~\ref{fig:concentrationMR}.
The vertical dashed line represents the canonical asymmetry threshold of $A\geq0.35$ \citep{Conselice03}, while the two dotted vertical lines represent arbitrary thresholds of $A\geq 0.8$ and 1.5 (\S\ref{sec:timedep}).
}
\label{fig:asymmetryMR}
\end{figure*}

Figure~\ref{fig:asymmetryMR} is constructed in a similar manner to
Figure~\ref{fig:concentrationMR} but for the asymmetry parameter.  In
contrast to the concentration, we find a strong correlation between
asymmetry $A$ and a galaxy's merger history, with the power-law
fits having positive slopes and fairly small errors indicating a
highly significant detection of a correlation.  However, the scatter
around the best fit is still fairly large ($\sim0.44$ dex), which
means that there is not necessarily a tight correlation between
$\cal{R}$ and $A$.  The correlation appears to be significantly
strengthened by a population of high-$\cal{R}$ objects that form
an upward spur at the highest $A$ values.

The canonical merger separation value of $A=0.35$ \citep{Conselice03}
is shown as the vertical dashed line, while higher values of $A=0.8$
and $A=1.5$ are shown as dotted lines that will be investigated in
\S\ref{sec:morphstatshist}.  Owing to the large scatter, there are
still many galaxies classified as major mergers with asymmetry
values $\geq0.35$ even though their merger ratios are consistent
with minor-mergers or in-situ growth (lower right quadrant).  Out
of all the data points within this sample, only $\simeq9\%$ with
$A\geq0.35$ are classified correctly as major mergers according to
their merger ratios extracted from the simulation.

Two avenues can immediately be seen for potentially improving the
fidelity of merger identification using $A$.  First, it is clear
that all but one of the true mergers lie at $A\gg 0.35$, hence we
can cut at a higher value of $A$ to obtain a higher fraction of
mergers.  Second, it is possible that $A$ identifies mergers well,
but perhaps in an earlier phase that is not associated with the
final coalescence, or even a later phase.  We will investigate these
avenues to improve merger characterisation using $A$ in
\S\ref{sec:morphstatshist}.

\subsection{$Gini$ and $M_{20}$}
\label{sec:gm20}

The $Gini$ coefficient originated as an economics statistic
representing the income distribution of a nation's residents.  A
value of zero means perfect equality, wherein a value of unity
represents perfect inequality.  When used in galaxy morphology determinations,
it quantifies the ``lumpiness" of the light distribution, with no prior
assumptions about its morphology \citep{Abraham03, Lotz04}.  In the
context of galaxy images, a $Gini$ coefficient of zero means that
the flux is evenly distributed among all the pixels, while a value
of one means that all of the flux is concentrated in a single pixel.

To compute the $Gini$ coefficient we first sort each galaxy's pixels
by flux in increasing order before applying the following equation
\citep[e.g.][]{Lotz04}:
\begin{equation}
Gini=\frac{1}{\bar{f}n(n-1)}\sum^n_i (2i-n-1)f_i.
\label{eq:gini}
\end{equation}
Here $f$ is the flux of each pixel, and $n$ is the number of pixels in each image.

\mtwenty is defined as the normalised second-order moment of the
brightest 20\% of a galaxy's flux.  This quantity is calculated
by again sorting the pixels from highest to lowest flux, then summing
over the brightest pixels until reaching 20\% of the total flux, and
normalising by its total second order moment (M$_{tot}$):
\begin{equation}
M_{20}\equiv \log\left(\frac{\sum_i M_i}{M_{tot}}\right), {\rm while}\, \sum_i f_i < 0.2f_{tot},
\end{equation}
where $f_{tot}$ is the total flux of the galaxy, and $f_i$ are the
fluxes in each pixel ordered from highest to lowest flux.  $M_{tot}$
is defined as the flux sum of each pixel multiplied by the squared
distance to the centre of the galaxy:
\begin{equation}
M_{tot}\equiv \sum_i^n M_i = \sum_i^n f_i \left[(x_i - x_c)^2 + (y_i - y_c)^2\right],
\end{equation}
where ($x_c,y_c$) is the galaxy's centre, and ($x_i ,y_i$) is the pixel location.

This value encodes similar information as the concentration parameter
(\S\ref{sec:ca}), but differs in two primary aspects: it is
more heavily weighted by the spatial distribution of luminous regions
because of the $r^2$ dependence, and it is not measured with circular
or elliptical apertures.  \citet{Lotz04} argues that these differences
make $M_{20}$ more sensitive to merger signatures than the concentration
parameter.

\begin{figure}
\includegraphics[width=0.5\textwidth]{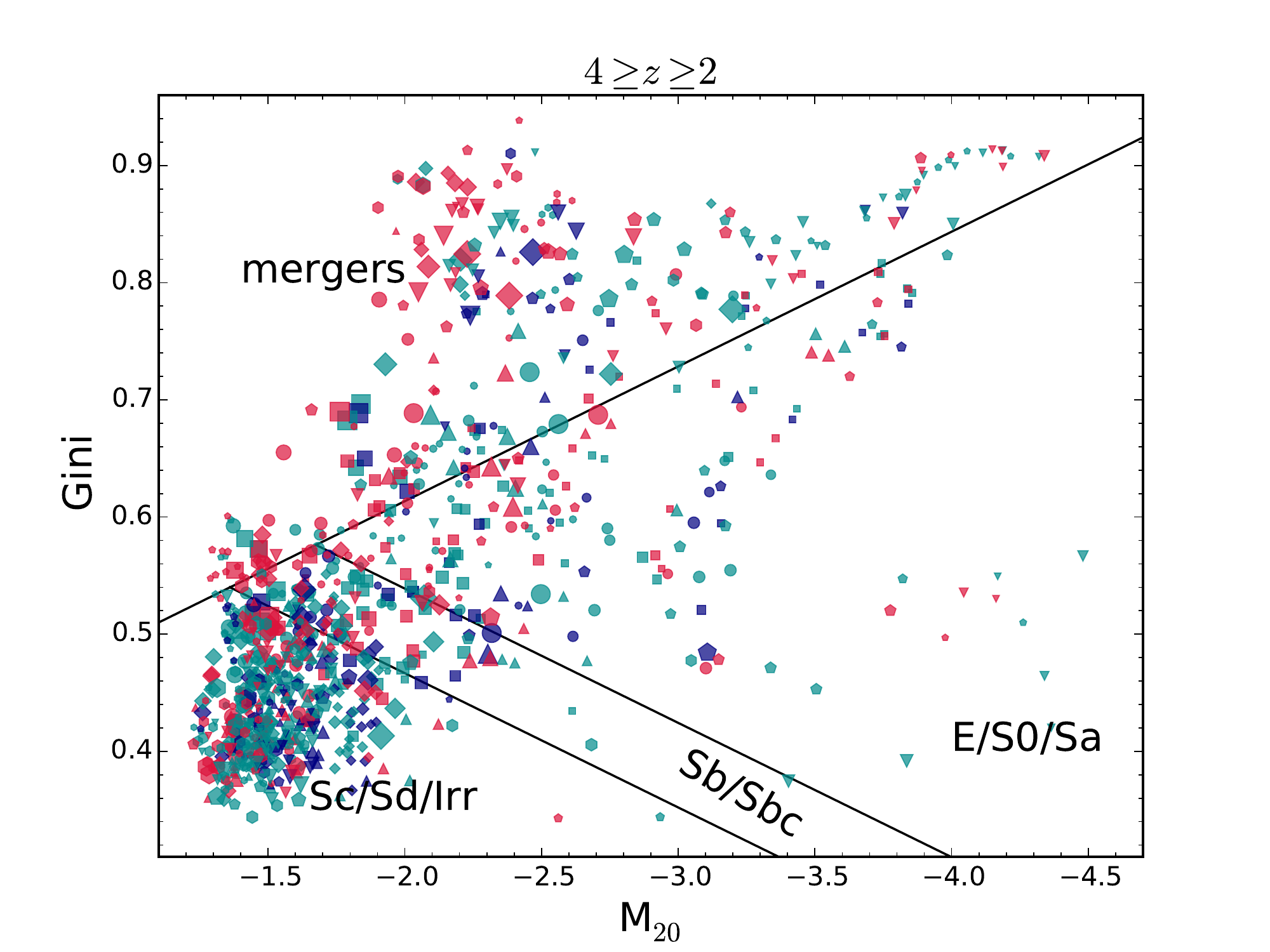}
\caption{
$Gini-M_{20}$ relation for our sample of galaxies plotted in the same fashion as Figure~\ref{fig:ca}.
The different regions on this plot represent the different morphological classifications that have been applied to local galaxies, and extrapolated to higher redshift observations.
}
\label{fig:gm20}
\end{figure}

Akin to $C$ and $A$, $G$ and \mtwenty are routinely plotted as a
function of one another which we show in Figure~\ref{fig:gm20} for
our sample.  The different regions on this plot represent the
different morphological classifications that have been applied to
local galaxies \citep{Lotz08}.  Within our sample, we find a median
$Gini$ coefficient of 0.51, and a median $M_{20}$ of -1.74.  Hence
our typical $z=2$ galaxy is an Sb/Sc spiral, which is
consistent with the images shown in Figure~\ref{fig:postage}.

\begin{figure*}
\includegraphics[width=1\textwidth]{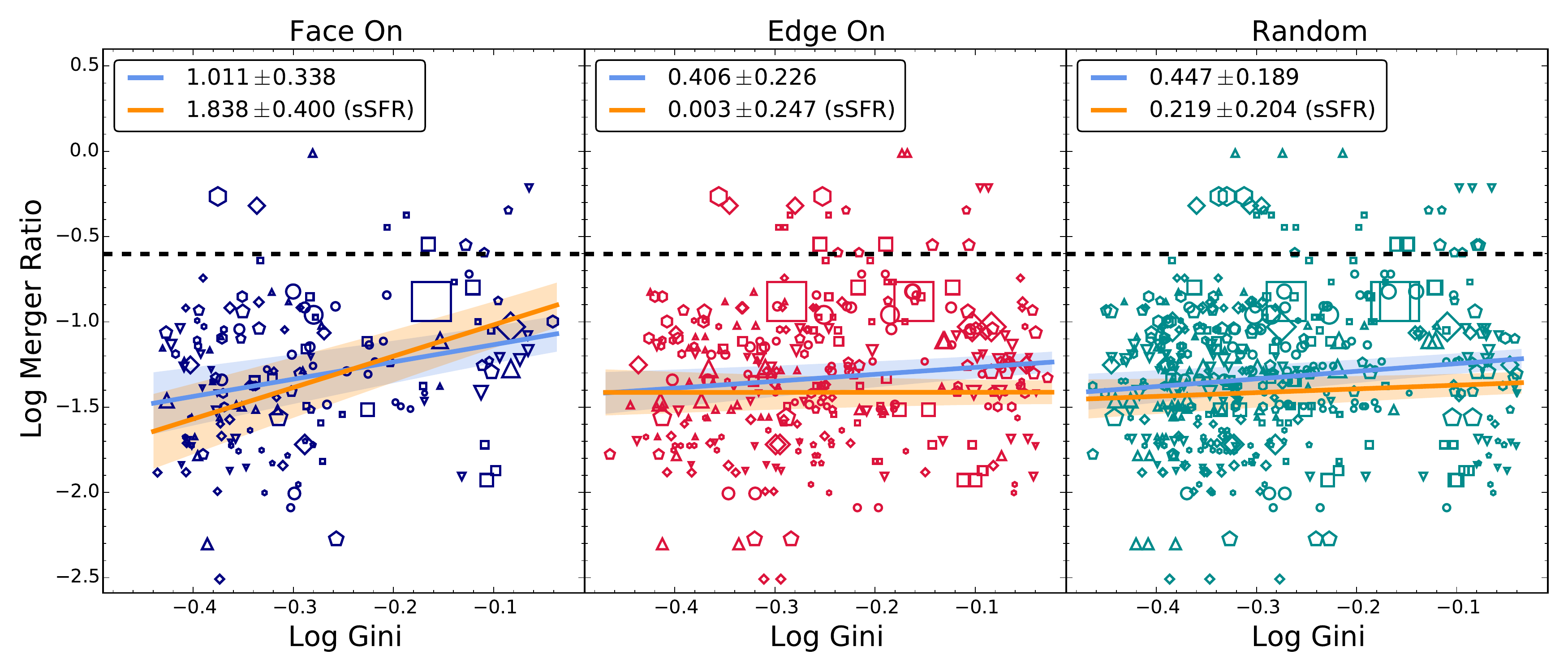}
\caption{
The merger-ratio $R$ as a function of $Gini$,
similar to Figure~\ref{fig:concentrationMR}.
}
\label{fig:giniMR}
\end{figure*}

Figure~\ref{fig:giniMR} plots the merger ratio ($\log \cal{R}$) versus
the Gini coefficient ($\log G$),
similar to Figure~\ref{fig:concentrationMR}.  By itself, Gini is
essentially uncorrelated with $\cal{R}$, except for a weak correlation
in the face-on orientation.  Not surprisingly, the Gini coefficient
alone offers a poor way to identify mergers.

\begin{figure*}
\includegraphics[width=1\textwidth]{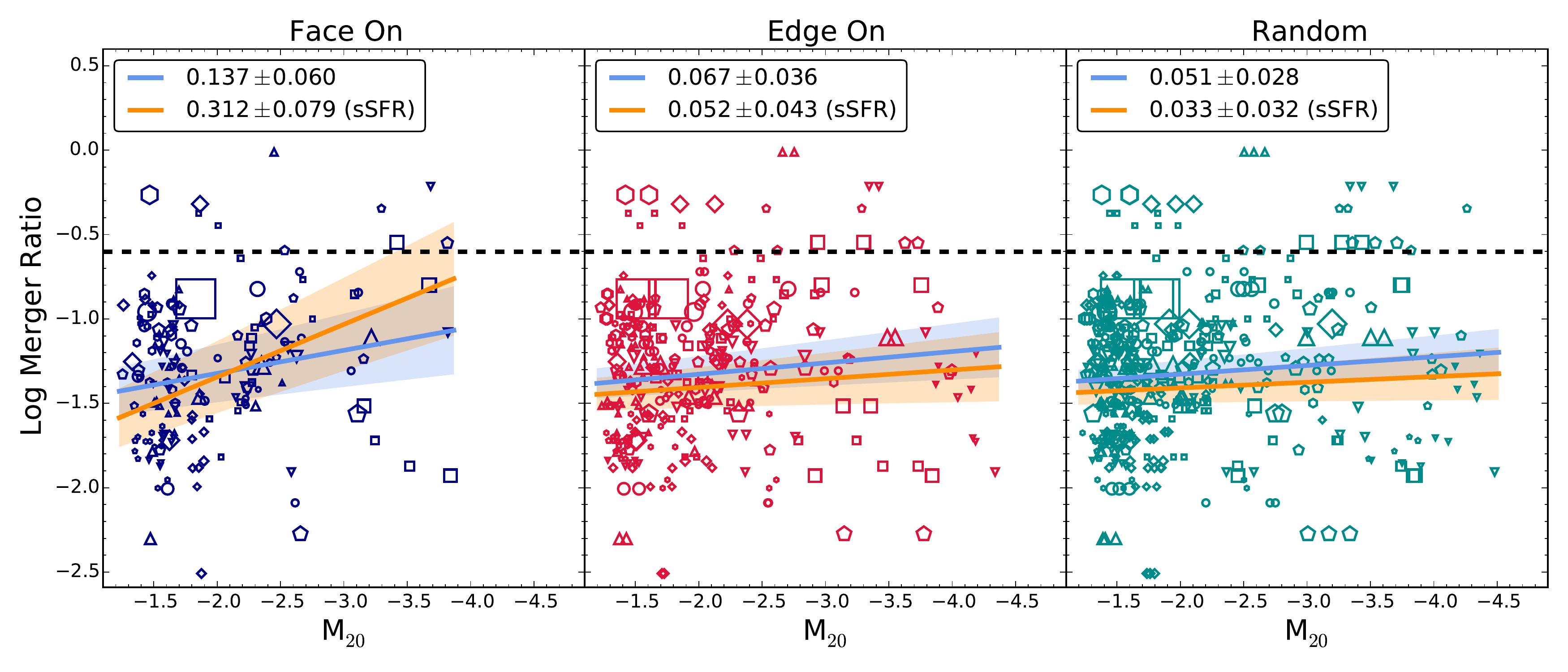}
\caption{
The merger-ratio $R$ as a function of $M_{20}$,
similar to Figure~\ref{fig:concentrationMR}.
}
\label{fig:m20MR}
\end{figure*}

Figure~\ref{fig:m20MR} plots the merger ratio as a function of
the $M_{20}$ parameter, 
similar to Figure~\ref{fig:concentrationMR}.
Again, there is only at most a weak anti-correlation of $M_{20}$
and merger ratio, and only in the face-on case.  Hence both Gini
and $M_{20}$ in isolation only can be associated with mergers in
the face-on case, and even then not reliably.  

We note that \citet{Stott13} used visual classifications in the
HiZELS survey at $z\sim 0.4$ to calibrate the relationship between
Gini, $M_{20}$, and disturbed morphologies indicative of mergers.
They found that a Gini-independent cut of $M_{20}>-1.5$ was efficient
at detecting visually-classified mergers.  Our results would not
support such a cut being used at $z\sim 2$,
as it would not do well at identifying galaxies that
have just undergone a merger.

\subsection{Merginess}
\label{sec:merginess}

While Gini and $M_{20}$ independently do not do well at identifying
mergers, typically merger classification relies on a combination
of these two parameters.  To quantify how well the \gm relation
classifies each data point as a merger, we define the term {\it
`merginess'} as the perpendicular distance to the canonical dividing
merger line \citep[similar to what was done in][]{Snyder15}.  A
point above this line (\gm merger) is assigned a positive value,
while a point below (\gm non-merger) is assigned a negative value,
given by the distance to the line.  The numerical value of this
distance in \gm space has no particular physical interpretation,
but can be used as a relative measure to identify how well the \gm
relation classifies mergers.

\begin{figure*}
\includegraphics[width=1\textwidth]{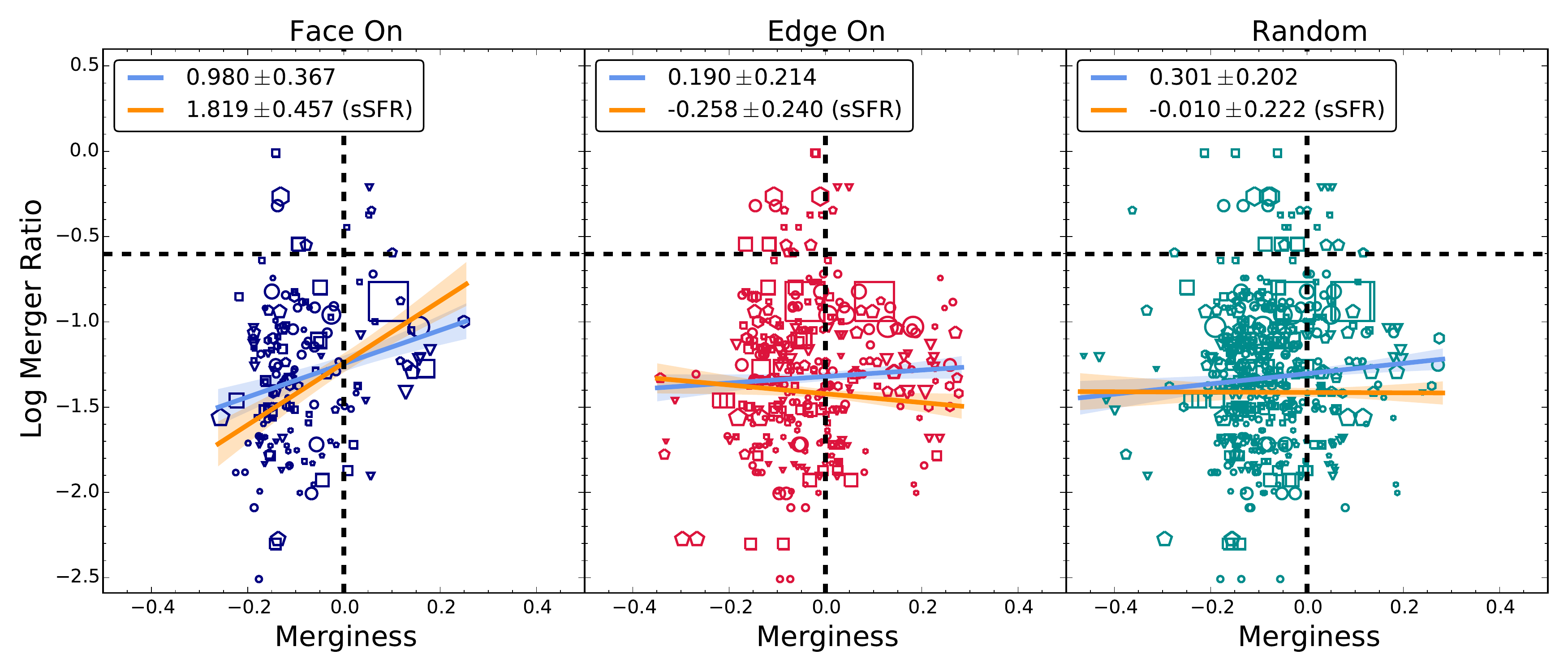}
\caption{ The merger-ratio $R$ as a function of the merginess,
similar to Figure~\ref{fig:concentrationMR}.
We additionally plot a vertical dotted line at merginess$=0$ to differentiate between a merger (positive merginess) and a non-merger (negative merginess) on the $Gini-M_{20}$ plane.
}
\label{fig:merginessMR}
\end{figure*}

Figure~\ref{fig:merginessMR} shows $\cal{R}$ of each object in our
sample as a function of merginess.  Every data point to the left
of the vertical dashed line is not classified as a merger on the
\gm plane, while everything to the right is a merger.  Each data point above
the dashed horizontal line ($R=-0.46$) represents a true major merger 
(between the previous two outputs) of at least a 1:4 mass
ratio.  Points in the upper right and lower left quadrants indicate
objects that are correctly classified; points in the other two
quadrants are incorrectly classified.

Figure~\ref{fig:merginessMR} indicates that the location in the \gm
relation does not do particularly well at identifying mergers at
these redshifts.  Points in the upper left quadrant are major mergers
in the simulation, yet are classified as non-mergers on the \gm
plane.  Points in the lower right on the other hand, have merger
ratios consistent with minor-mergers yet are classified as major mergers 
on the \gm plane.  Out of the 251 data points classified
as a merger in the \gm relation (positive `merginess'), only 18
($\sim7\%$) lie above the ${\cal R}>0.25$ line and are a direct
result of major mergers.

Fits to the data suggest very little correlation, with the
exception of face-on galaxies that have a power-law slope that
deviates by more than $3\sigma$ from zero.  This correlation is
still much less significant than for the asymmetry ($A$).  In particular,
unlike for $A$ where many of the true mergers lie at high-$A$, here
the true mergers are not well-localised in merginess.  Even for the
face-on case, there are only 5 major mergers out of the 31 with
positive merginess ($\sim 16 \%$) that are correctly classified.

In summary, of the non-parametric morphological statistics examined
here, only asymmetry provides a clear correlation with mergers,
with the \gm-based merginess parameter showing some correlation
only in the case of face-on galaxies.  Recall, however, that our
merger quantification only picks out major mergers that have occurred
between the previous two outputs.  It is possible, as was mentioned
for asymmetry, that these statistics might not identify mergers
exactly when they end, but perhaps they do better at identifying
recent or future mergers.  We examine this in the next section.


\section{Morphological statistics in time}
\label{sec:morphstatshist}

In this section we plot each morphological statistic and
merger ratio over varying intervals in cosmic time.  First, we
qualitatively examine the evolution of each of our statistics, in
relation to the evolution of the merger ratio.  Then, we 
quantify how well each statistic does at tracking mergers over a
range of time intervals.  For brevity, we omit consideration
of Gini and $M_{20}$ individually, and focus only on the merginess
parameter defined by the location in \gm space.


\subsection{Evolution of statistics}
\label{sec:concentrationhist}

In the figures of this section, we overlay the evolution of the
merger ratio $\cal{R}$ and each morphological statistic from $4\geq
z\geq 2$.  We consider separately the three cases of face-on,
edge-on, and random orientations; for the two cases with multiple
data points (edge-on and random) we take the median value at any
given redshift.  We plot the  merger ratio $\cal{R}$ as a
dashed line and plot the dividing line between major and minor
mergers as a horizontal dotted line; hence major mergers are when
the dashed line goes above the dotted line.

\begin{figure*}
\includegraphics[width=1.1\textwidth]{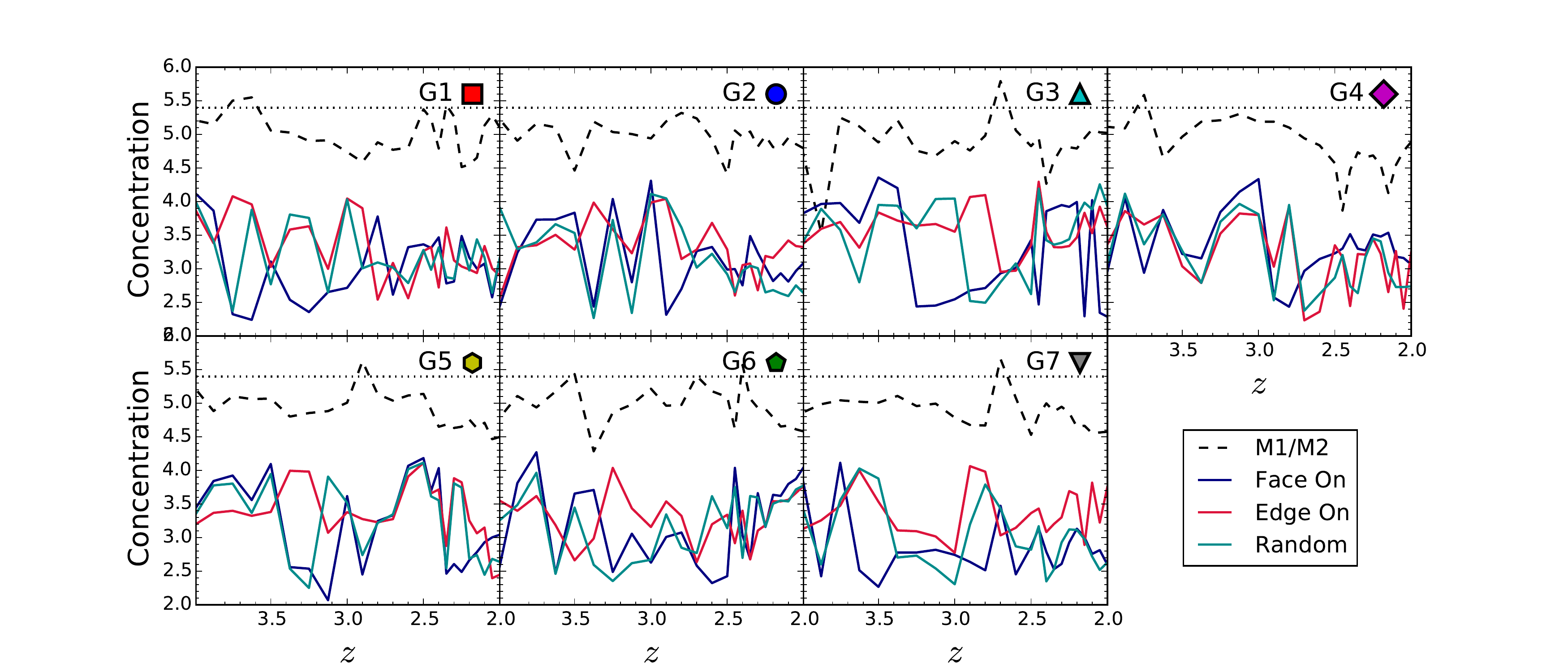}
\caption{
The concentration parameter $C$ as a function of cosmic time for our seven galaxies.
We consolidate our 6 viewing angles into either face-on, edge-on, or random; because the latter two have multiple data points we take the median values.
We also show the normalised merger-ratio history of each galaxy as the dashed line, with the dotted line representing the threshold for a major merger; a major
merger occurs when the dashed line crosses above this dotted line.
}
\label{fig:concentrationhist}
\end{figure*}

Figure~\ref{fig:concentrationhist} shows the concentration of each
of our seven galaxies in various orientations as a function of
cosmic time.  We see that there is no indication that $C$ is strongly
impacted by major mergers, in accord with our findings in \S\ref{sec:ca}.
In some instances the concentration parameter will spike shortly
before or after a major merger, but overall there is no consistent
correlation between a high $\cal{R}$ and $C$.  This is not surprising
since $C$ is not typically used to classify mergers.

\begin{figure*}
\includegraphics[width=1.1\textwidth]{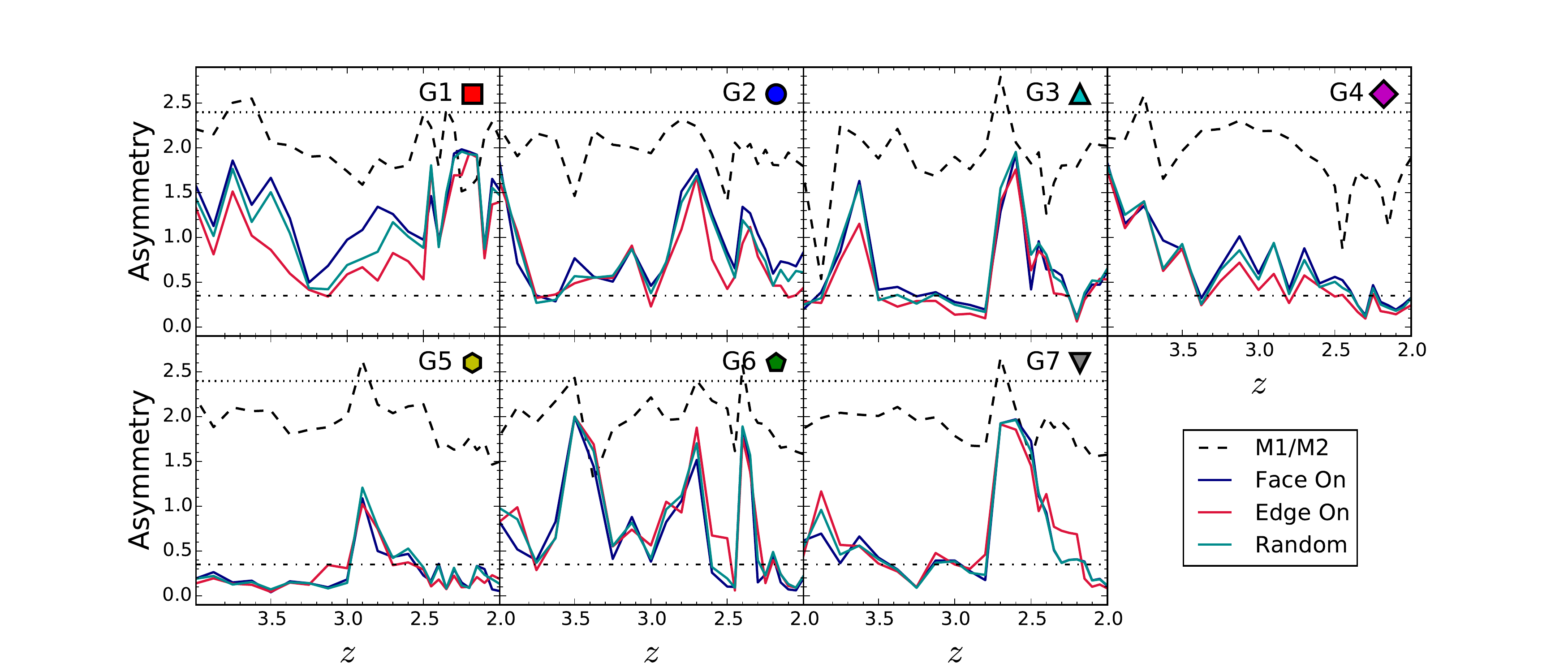}
\caption{
The asymmetry parameter $A$ as a function of cosmic time for our seven galaxies, similar to Figure~\ref{fig:concentrationhist}.
Asymmetry values above $A\geq0.35$ are canonically considered major mergers \citep{Conselice03}; we plot this threshold as a dot-dashed line.
}
\label{fig:asymmetryhist}
\end{figure*}

Figure~\ref{fig:asymmetryhist} similarly shows the asymmetry $A$ as a
function of time, with $\cal{R}$ overlaid as before.  Here we
immediately notice that in many cases, spikes in $A$ mimic spikes
in $\cal{R}$.  The correlation is certainly not tight, but it does
appear that overall when $\cal{R}$ is large, $A$ is often large
as well.  Again, this reflects the results we obtained in \S\ref{sec:ca},
but with the additional information that there is a more general
temporal correlation between these quantities.

We argued that a higher cut than the canonical $A\geq0.35$
\citep[indicated by the horizontal dot-dashed
line;][]{Conselice03} may be more effective at identifying
mergers.  This is also evident in this plot, but the optimal cut
is not obvious; each galaxy would require a different cut in $A$
to allow for a complete selection of major mergers.  For G1, G3,
G4, G6, and G7 for instance, a cut around $A\ga2$ would identify
mergers well.  But this cut would not work for G2 since it never encounters
a major merger yet the asymmetry approaches two at $z\sim2.7$.  For
G5, conversely, its asymmetry history stays below the
\citet{Conselice03} $A\geq0.35$ cut until the major
merger event, at which point $A$ exceeds 0.35 and remains above the
cut for $\sim3$ outputs after the major merger.  This particular
galaxy has a similarly smooth growth history similar to G2, yet its
asymmetry history is much calmer.  G5 is the galaxy within our
sample showing the most well-settled disk-like morphology at $z=2$
akin to low-redshift galaxies, which may help to explain why a cut
of $A\geq0.35$ can correctly identify the major merger for this
galaxy alone.

\begin{figure*}
\includegraphics[width=1.1\textwidth]{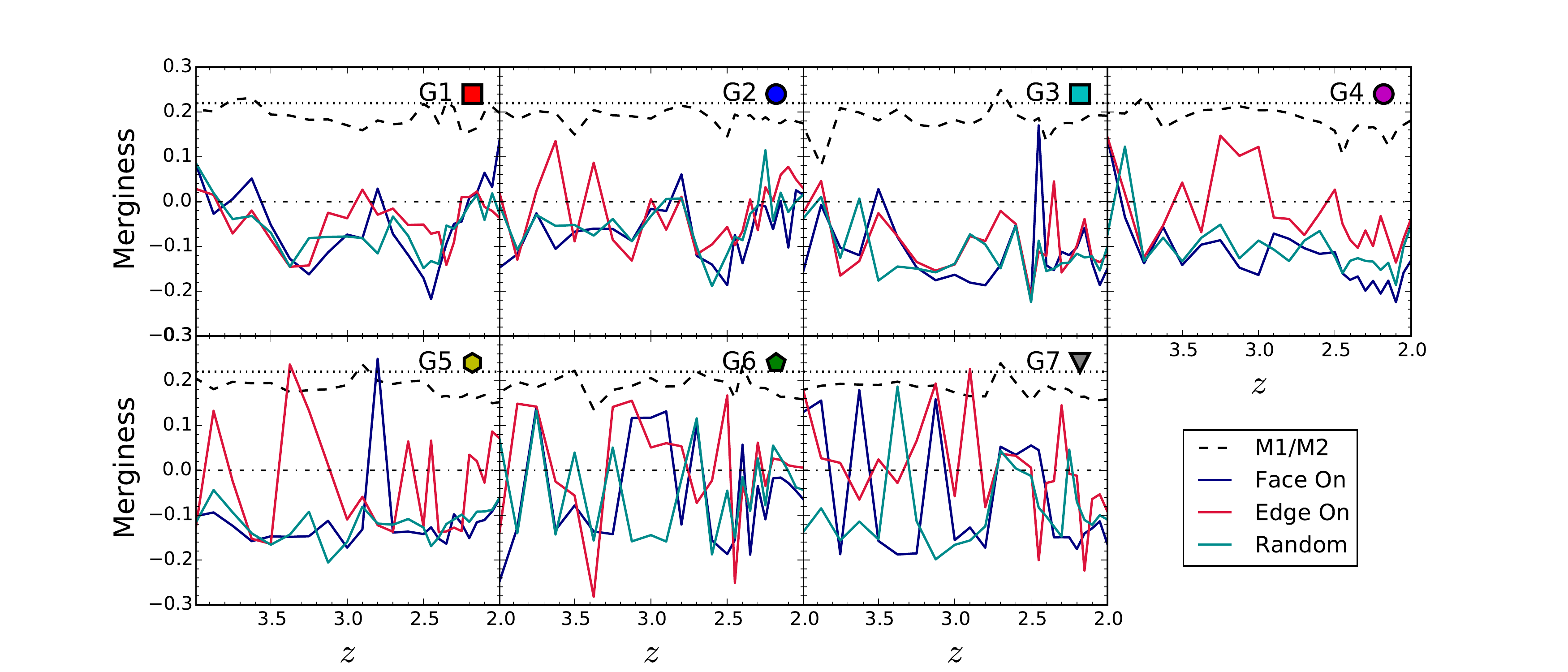}
\caption{
The `merginess' as a function of cosmic time for our seven galaxies, similar to Figure~\ref{fig:concentrationhist}.
The dot-dashed line represents the difference between a merger (positive merginess) and a non-merger (negative merginess) in the $Gini-M_{20}$ plane.
}
\label{fig:merginesshist}
\end{figure*}

Figure~\ref{fig:merginesshist} analogously shows the merginess as
a function of cosmic time for the galaxies in our sample.  It is
clear that the actual merger event does not straightforwardly
increase the merginess, as we saw in \S\ref{sec:merginess}.  On
occasion, however, the merginess spikes at a time that is close to,
but not exactly coincident with, the merger event.  We will quantify
this in the next section.

\subsection{A quantitative assessment of {\bf $A$} and merginess in time}
\label{sec:timedep}

We now ask the question, do $A$ and merginess identify mergers that
have happened within a certain time interval from the time of
observation?  In other words, if one observes a galaxy at a given
redshift to have a high value of $A$ or merginess, what is the
probability that it recently underwent or will soon complete a
merger?

To quantify this, we re-assign the values of $\cal{R}$ around our
ten major merger events for $\pm N$ surrounding time steps to the
$\cal{R}$ at the time of the major merger itself.  This equates to
adjusting the $y$-axis values in
Figures~\ref{fig:asymmetryMR}~and~\ref{fig:merginessMR} for data
points $\pm N$ time steps around each major merger, guaranteeing
that these data points will be considered a major merger according
to their merger ratios.  We can then measure the fraction of objects
above a given asymmetry or merginess threshold that also have a
$\cal{R}$ consistent with a major merger.  In this way, we can
quantitatively answer the question, does a high value of $A$ or
merginess correspond to identifying a merger within some time
interval $\Delta t$ (corresponding to $\pm N$ outputs)?

\begin{figure*}
\includegraphics[width=1.1\textwidth]{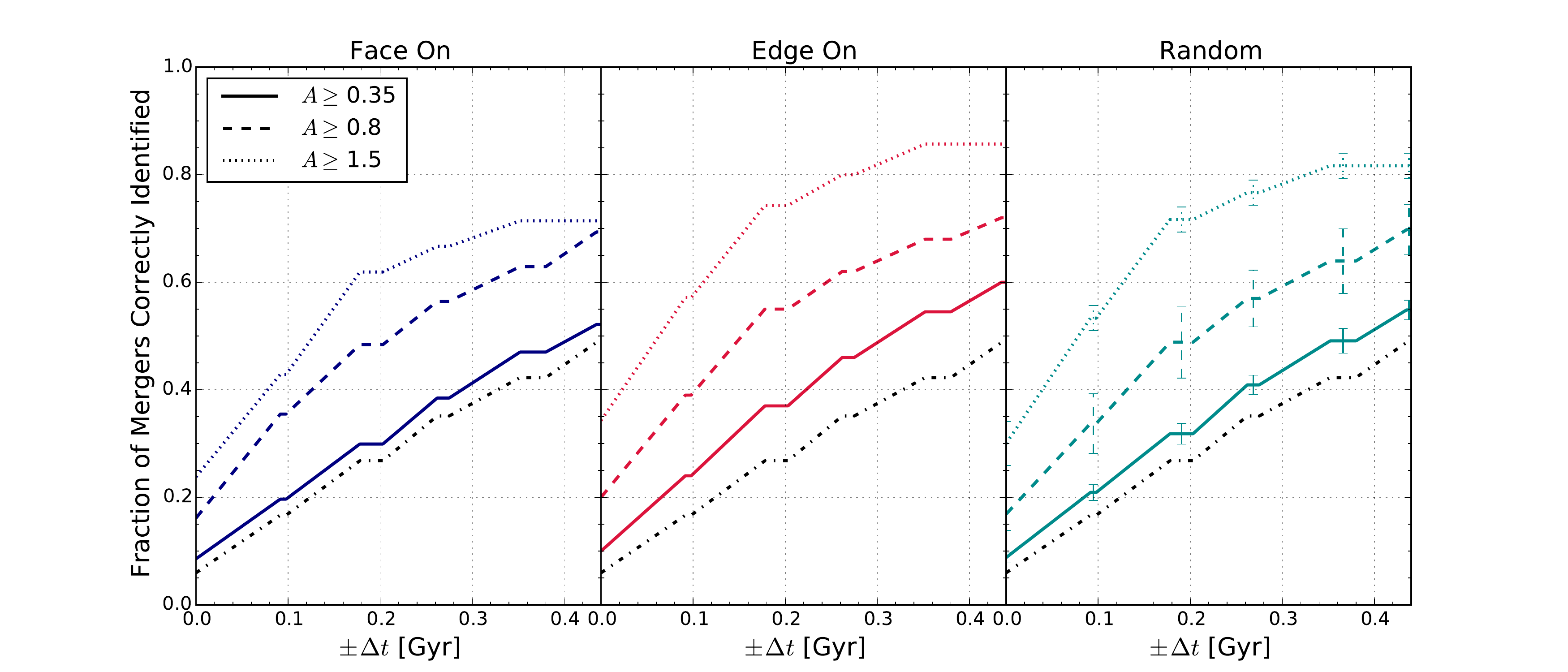}
\caption{
The fraction of major mergers correctly identified above a given asymmetry threshold when we consider the time dependent effects discussed in \S\ref{sec:timedep}.
The solid lines represent the typical asymmetry cut \citep[$A\geq0.35$;][]{Conselice03} used to identify major merger candidates, while higher cuts of 0.8 (dashed) and 1.5 (dotted) are arbitrary choices to illustrate the effectiveness of $A$ at selecting major mergers within our sample.
The black dot-dashed line is the chance of encountering a random merger over a given $\pm \Delta t$ as if there were no threshold value (background noise).
Error bars represent the spread in the fraction mergers correctly identified when examining the $x$, $y$, and $z$ viewing orientations independently.
}
\label{fig:correctAsymmetryFraction}
\end{figure*}

Figure~\ref{fig:correctAsymmetryFraction} shows these fractions for
various thresholds of $A$ for our different viewing orientations,
as a function of the time interval $\pm\Delta t$. 
The solid lines represent the fraction of
correctly identified major mergers for a given $\Delta t$ with
$A\geq0.35$ \citep{Conselice03}, while the dashed and
dotted lines show $A\geq 0.8$ and $A\geq 1.5$, respectively.
The dot-dashed line considers all values of $A$ (i.e. $A>0$),
so effectively shows the ``background noise" that signifies the
likelihood of being within $\pm \Delta t$ of a merger at any
given time.
Error bars for the random viewing orientation represent the
spread in the fraction of correctly identified mergers for
the $x$, $y$, and $z$ viewing orientations independently.

We can see that a galaxy within our sample having $A\geq0.35$
provides little better merger identification than the noise, for
any time interval and viewing orientation.
However, increasing the cut to $A\geq 0.8$
significantly increases the ability to identify mergers.  A galaxy
satisfying this cut has a $\sim50\%$ chance of being within $\pm 200$~Myr
of a merger.  $A\geq 1.5$ provides even higher percentages, so that
there is now a $\sim70\%$ chance of a major merger within $\pm
0.2\rm{Gyr}$.  The percentages are somewhat higher for edge-on
systems than face-on, indicating that $A$ is better at identifying
mergers in edge-on systems.

There is also the converse question, i.e. what fraction of true
mergers are identified as such by these cuts?  For $A\geq0.35$, it
finds all but one of the true mergers (while also unfortunately finding many
that are not).  $A\geq0.8$ identifies 90\% (9/10) of all major mergers,
while $A\geq1.5$ identifies only 60\% (6/10) of them.  Hence, while an
increased cut improves the accuracy of merger identification, it
sacrifices the ability to identify a highly complete sample of merger
events.

\begin{figure*}
\includegraphics[width=1.1\textwidth]{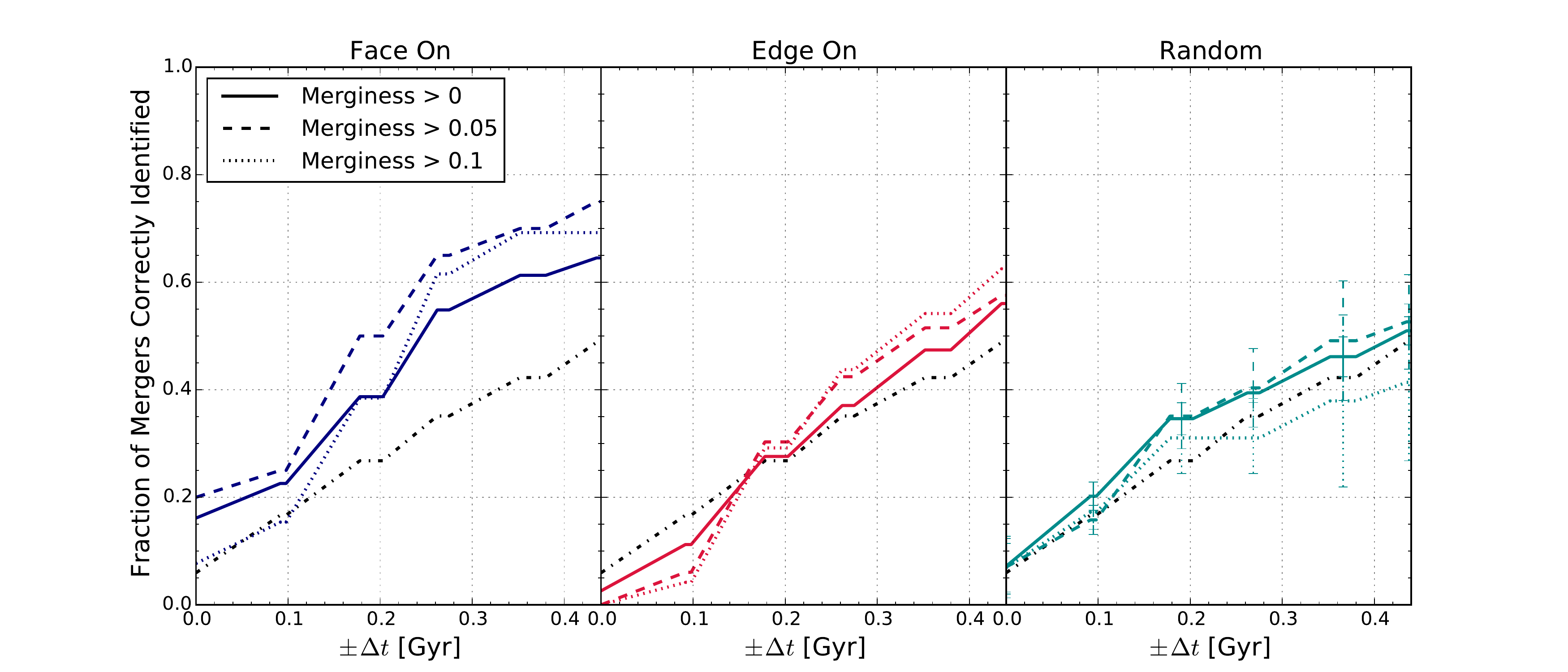}
\caption{
The fraction of major mergers correctly identified above a given merginess threshold when we consider the time dependent effects discussed in \S\ref{sec:timedep}.  
The solid lines represents a merginess threshold of merginess=0 used to identify major merger candidates on the \gm plane, while higher thresholds of 0.05 (dashed) and 0.1 (dotted) are arbitrary choices to illustrate the effect of shifting the merginess threshold.
The black dot-dashed line is the chance of encountering a random merger over a given $\pm \Delta t$ as if there were no merginess threshold (background noise).
Error bars represent the spread in the fraction mergers correctly identified when examining the $x$, $y$, and $z$ viewing orientations independently.
}
\label{fig:correctmerginessFraction}
\end{figure*}

Figure~\ref{fig:correctmerginessFraction} shows the analogous
fractions of correctly-identified mergers for various merginess
thresholds, as a function of $\pm\Delta t$.  For the random orientation
case, merginess provides essentially no benefit for identifying
mergers relative to the background noise.  Only in the face-on case
does merginess provide additional information regarding a merger,
and interestingly only for $\Delta t\ga 200$~Myr.  Changing the
merginess cut does little to help, unlike in the asymmetry case
where a more stringent cut greatly improves the fidelity of merger
identification.

\begin{figure*}
\includegraphics[width=1.1\textwidth]{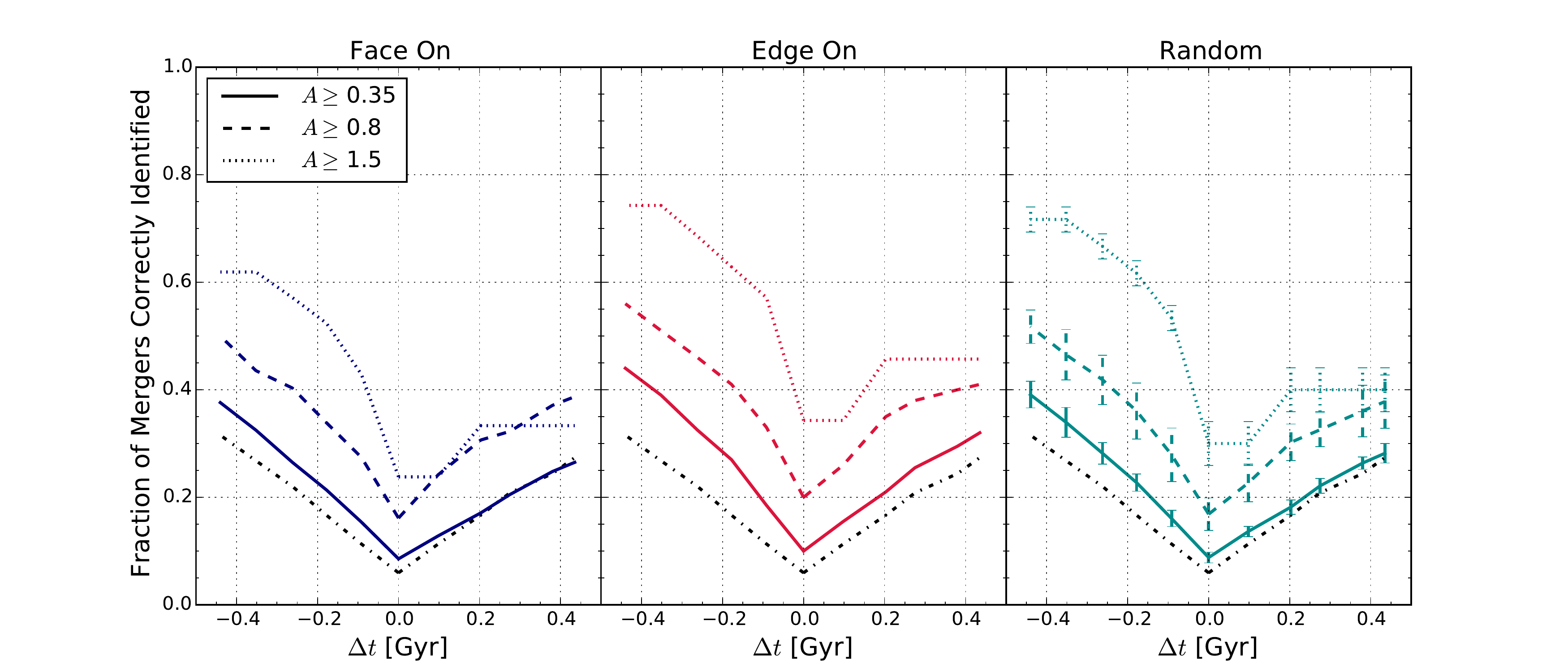}
\caption{
Similar to Figure~\ref{fig:correctAsymmetryFraction}, but here we plot the fraction of mergers correctly identified when separately considering the $+\Delta t$ and $-\Delta t$ time dependent effects discussed in \S\ref{sec:timedep} for the asymmetry parameter.
}
\label{fig:correctAsymmetryFractionBothDirs}
\end{figure*}

So far we have considered intervals symmetric in time, i.e. we have
not distinguished whether a given statistic better identifies
pre-mergers or merger remnants.
Figure~\ref{fig:correctAsymmetryFractionBothDirs} shows the
effectiveness of merger identification for $A$ considering positive
and negative $\Delta t$ separately.  It is clear that $A$ preferentially
identifies pre-coalescence mergers, and shows the same slight
preference for identifying these in edge-on systems as before.

\begin{figure*}
\includegraphics[width=1.1\textwidth]{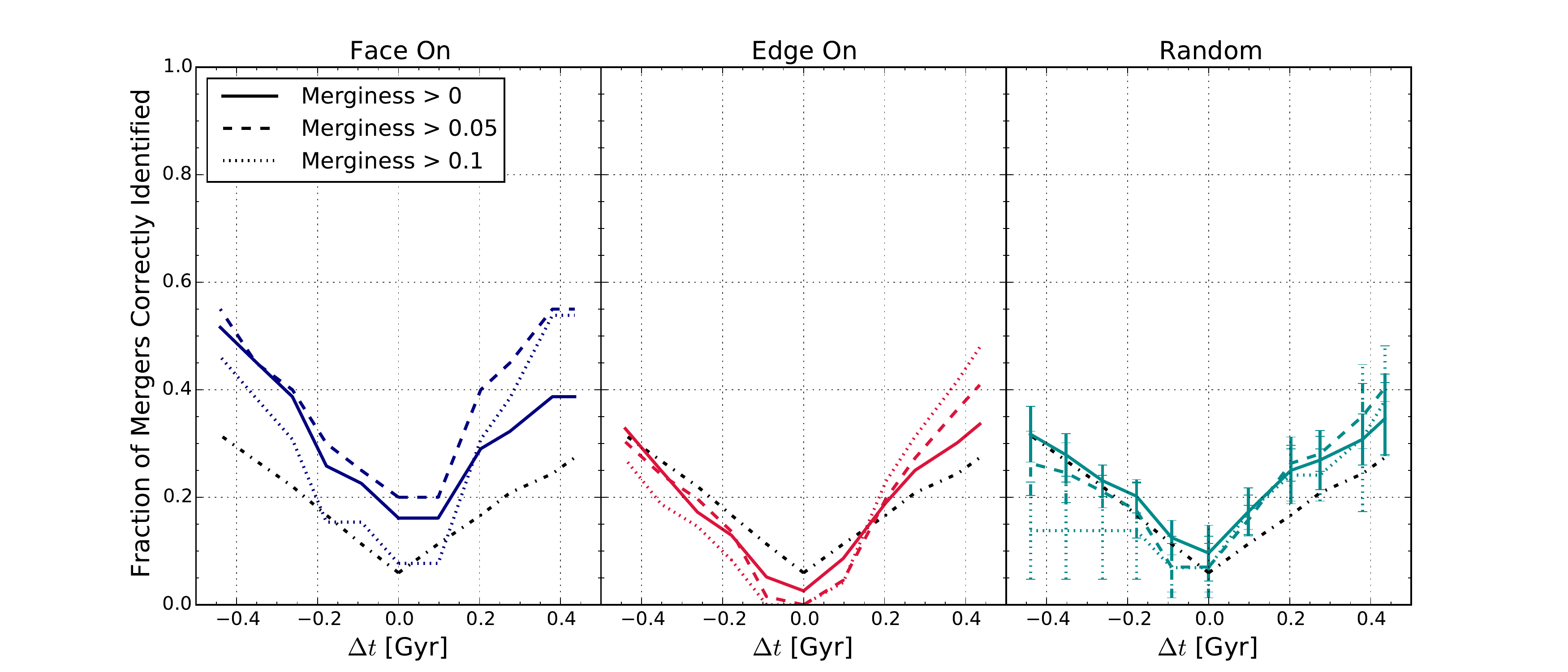}
\caption{
Similar to Figure~\ref{fig:correctmerginessFraction}, but here we plot the fraction of mergers correctly identified when separately considering the $+\Delta t$ and $-\Delta t$ time dependent effects discussed in \S\ref{sec:timedep} for the merginess parameter.
}
\label{fig:correctmerginessFractionBothDirs}
\end{figure*}

Merginess, on the other hand, displays a minor preference for
identifying merger remnants in the face-on case, as illustrated in
Figure~\ref{fig:correctmerginessFractionBothDirs}, but 
edge-on and random viewing orientations still provide little to no
benefit compared to the background noise.  Increasing the
merginess threshold typically yields a decrease in the fraction of
correctly identified mergers when looking in the $-\Delta t$
direction.  In the $+\Delta t$ direction, however, we find that
increases in the merginess threshold provides minor benefits only
at $\Delta t \ga 200 {\rm Myr}$. 

In short, a high value in the asymmetry ($A$), such as 0.8, provides the best
way to identify major mergers at $z\sim 2-4$.  It fares slightly better
in the edge-on systems than in face-on ones, but even for a random orientation
a galaxy with $A\geq0.8$ has a $\approx 50\%$ chance of being
involved in a merger within $\pm 0.2$~Gyr, and it identifies all but
one merger in our sample.  A higher cut of $A>1.5$ provides a greater
probability for being (temporally) close to a merger, at the cost
of a significantly incomplete sample of mergers.  The location in the
\gm plane, as quantified by `merginess', is only mildly effective
at identifying mergers in the face-on case, and ineffective in the
edge-on or random orientations.  Moreover, no other cut in merginess
is significantly more effective.  We thus conclude that asymmetry
represents a better, though not perfect, approach to identifying
mergers in high-$z$ galaxies than \gm.

\subsection{Resolution dependence}
To assess the resolution dependence of our results,
we examine the lower-resolution counterparts to the 
galaxies listed in Table~\ref{table:sims} from the 
original low-resolution full box simulation (\S\ref{sec:ics}).
By subtracting the fractional value from the background noise (\S\ref{sec:timedep})
we can quantify the increase in the fraction of mergers correctly 
identified in relation to the background noise for each of our $A$ 
and merginess cuts.

\begin{figure}
\includegraphics[width=0.52\textwidth]{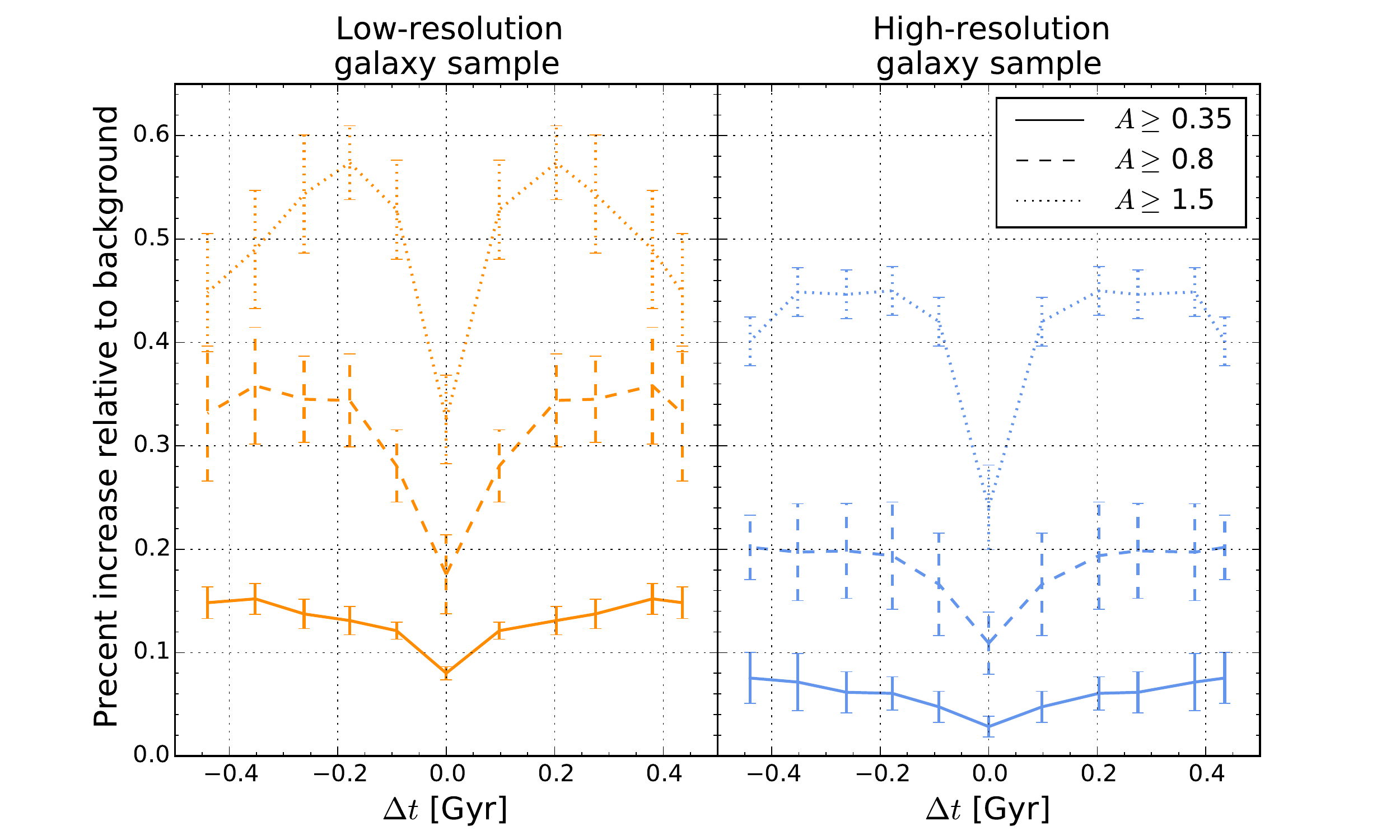}
\caption{
Fractional increase in the fraction of correctly identified major mergers
when compared to the background noise 
(\S\ref{sec:timedep}) for three cuts in $A$ and random viewing orientation.
{\it Left:} results from the low-resolution galaxy sample (\S\ref{sec:ics}),
{\it Right:} results from the high-resolution zoom sample used throughout this work.
Error bars represent the spread in the increase in the fraction of correctly identified
mergers when examining the $x$, $y$, and $z$ viewing orientations independently.
}
\label{fig:asymmetryZC}
\end{figure}

\begin{figure}
\includegraphics[width=0.52\textwidth]{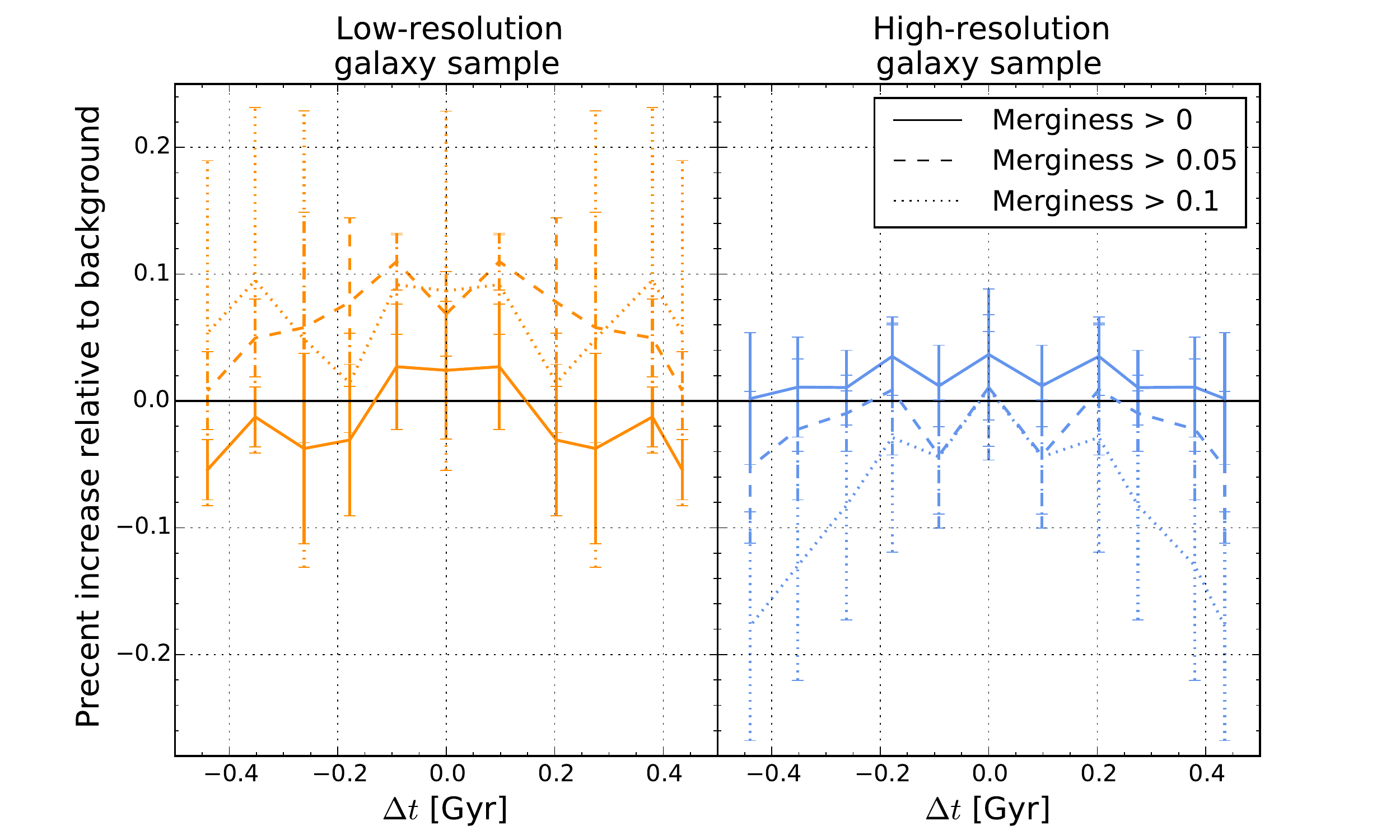}
\caption{
Similar to Figure~\ref{fig:asymmetryZC} for the merginess parameter.
}
\label{fig:merginessZC}
\end{figure}

Figures~\ref{fig:asymmetryZC} and \ref{fig:merginessZC} show this
fractional increase with respect to the background noise for the random viewing orientation
(we omit the face-on and edge-on viewing orientations as they show similar trends).
The left panels show the results for the low-resolution galaxy sample, 
while the right panels shows the high-resolution zoom sample.
As in Figures~\ref{fig:correctAsymmetryFraction}-\ref{fig:correctmerginessFractionBothDirs}, 
the error bars represent the spread in the percentage increase 
in the fraction of correctly identified mergers when examining the 
$x$, $y$, and $z$ viewing orientations independently.

We find that the general conclusions drawn here as regards the 
efficacy of $A$ and merginess are not overly sensitive to resolution.
We do find that the overall fraction of correctly identified systems along
with the background noise level both increase somewhat with a decrease in resolution, 
but the difference between the two remains similar.

\section{Conclusions}
\label{sec:conclusions}

We have performed high-resolution cosmological zoom-in simulations
to investigate the evolution of nonparametric galaxy morphological
statistics and their relationship to major mergers at $z\geq2$.  We
consider 7 galaxies with $3.5\times 10^{9}<M_*<6.6\times 10^{10} M_\odot$
and $2\times 10^{11}<M_{\rm halo}<10^{12} M_\odot$
at $z=2$ having star formation rates from $5-60 M_\odot$yr$^{-1}$.
By post-processing our simulated galaxies through {\sc Loser}, we
create mock images of each object as if it were observed with {\it HST}
WFC3's F160W filter.  The fluxes from our simulated images are used
to calculate the concentration ($C$), asymmetry ($A$), $Gini$
coefficient ($G$), and the normalised second-order moment of the
brightest 20\% of a galaxy's flux ($M_{20}$).  We track the mass
growth history of each galaxy from $4\geq z \geq 2$ and identify
10 major merger events of at least a 1:4 mass-ratio, and search for
correlations with the above mentioned morphological statistics that
could most reliably identify such merger events.

Our main findings are as follows:

\begin{itemize}

\item The asymmetry statistic ($A$) strongly correlates with merger ratio
among our simulated galaxies at these redshifts
(Figures~\ref{fig:asymmetryMR} and \ref{fig:asymmetryhist}).
Concentration, $Gini$, and $M_{20}$ do not substantially correlate
with merger ratio.  We define a `merginess' parameter that quantifies
the distance from the merger dividing line in \gm space, but this
also does not correlate with merger ratio, except for a weak
correlation in the face-on case (\S\ref{sec:merginess}).

\item Examining $A$ in more detail, we find that a higher threshold
for $A$ than the canonically-used 0.35 provides a more robust
identification of mergers.  For instance $A\geq0.8$ identifies 9/10
of our mergers, with a much lower fraction of erroneously-identified
mergers, $\sim 50$\%.  $A\geq1.5$ further reduces erroneous identifications, but
only identifies about half the mergers in our sample.

\item We examine the temporal dependence of these parameters by
quantifying how well each statistic identifies a merger that is
about to happen or has just happened within some time $\Delta t$
(\S\ref{sec:timedep}).  The merginess parameter (from \gm)
only weakly identifies mergers above random sampling in the face-on
case, but does not properly identify mergers in the random orientation
case.  (Figure~\ref{fig:giniMR}).

\item The probability of $A$ correctly identifying a merger increases
with both $\Delta t$ and the threshold value.  A galaxy with
$A\geq0.8$ has a $\sim 50\%$ chance of being an object that experiences
a merger within $\pm 200$~Myr.  The probability increases to
$\sim70\%$ for $A\geq1.5$, at the cost of a less complete sample
of mergers.  (Figure~\ref{fig:correctAsymmetryFraction}).  $A$ also
preferentially identifies pre-coalescence mergers
(Figure~\ref{fig:correctAsymmetryFractionBothDirs}).

\item Within our sample, compact galaxies \citep[as defined
by][]{Barro13} are formed naturally via in-situ growth or minor-mergers.
In some instances major mergers can briefly force a galaxy to become
compact, after a period where galaxies become more extended as
measured by their half-light radius.

\end{itemize}

Overall, we find that the best approach for identifying mergers
from $z=2$ to 4 is to use the asymmetry statistic ($A$) with a higher
cut than used for low-$z$ galaxies.  With larger samples we will
be able to quantify these statistics in more detail, and in future
work we will also examine the sensitivity to numerical methodologies
in greater detail.  It is also possible that resolved colour maps
may provide additional information regarding the evolutionary state
of galaxies.  It is hoped that these types of investigations will
aid the interpretation of resolved studies of high-redshift galaxies
that have emerged using {\it HST} and will be significantly
furthered in the {\it JWST} era.


\section*{Acknowledgements}

Analysis performed with pyGadgetReader \citep{pygr} and SPHGR
\citep{SPHGR}.  Simulations were run on the University of Arizona's
El Gato supercomputer\footnote{\url{http://elgato.arizona.edu/}}
funded by NSF MRI grant 1228509, and the University of the Western
Cape's Timon and Pumbaa clusters.  This work was supported by NASA
grant NNX12AH86G.  RD and RT acknowledge support from the South
African Research Chairs Initiative and the South African National
Research Foundation.  This publication is funded in part by the Gordon and 
Betty Moore Foundation's Data-Driven Discovery Initiative through Grant GBMF4561.
The authors thank J. Lotz, G. Snyder, and R.
Somerville for helpful discussions.


\bibliography{references}

\end{document}